\documentclass[preprint,aps,12pt,showpacs,nofootinbib,tightenlines,amsmath,amssymb]{revtex4}

\usepackage{amsmath}
\usepackage{graphicx}
\usepackage{amssymb}
\newcommand{\Tr}{\ensuremath{\mathop{\mathrm{Tr}}}}
\newcommand{\<}{\langle}
\renewcommand{\>}{\rangle}
\newcommand{\ti}{{t_{\mathrm{i}}}}
\newcommand{\tf}{{t_{\mathrm{f}}}}
\newcommand{\x}{{\vec x}}
\renewcommand{\P}{{\vec P}}

\begin{document}
\def\intdk{\int\frac{d^4k}{(2\pi)^4}}
\def\sla{\hspace{-0.17cm}\slash}
\hfill

\title{Chiral Thermodynamic Model of QCD and its Critical Behavior in the Closed-Time-Path Green Function Approach}

\author{Da Huang}\email{dahuang@itp.ac.cn} \author{ Yue-Liang Wu }\email{ylwu@itp.ac.cn}
\affiliation{ State Key Laboratory of Theoretical Physics (SKLTP) \\ 
Kavli Institute for Theoretical Physics China (KITPC) \\ 
Institute of Theoretical Physics, Chinese Academy of Sciences, Beijing,100190,
P.R.China}
\date{\today}

\begin{abstract}
By applying the closed-time-path Green function formalism to the
chiral dynamical model based on an effective Lagrangian of chiral
quarks with the nonlinear-realized meson fields as bosonized
auxiliary fields, we then arrive at a chiral thermodynamic model for
the meson fields after integrating out the quark fields. Particular attention is
paid to the spontaneous chiral symmetry breaking and restoration
from the dynamically generated effective composite Higgs potential
of meson fields at finite temperature. It is shown that the minimal
condition of the effective composite Higgs potential of meson fields
leads to the thermodynamic gap equation at finite temperature, which
enables us to investigate the critical behavior of the effective
chiral thermodynamical model and to explore the QCD phase
transition. After fixing the free parameters in the effective chiral
Lagrangian at low energies with zero temperature, we determine the
critical temperature of the chiral symmetry restoration and present
a consistent prediction for the thermodynamical behavior of several
physically interesting quantities, which include the vacuum
expectation value $v_o(T)$, quark condensate $<\bar{q}q>(T)$, pion
decay constant $f_\pi(T)$ and pion meson mass $m_{\pi}(T)$.
In particular, it is shown that the thermodynamic scaling behavior
of these quantities becomes the same near the critical point of phase transition.
\end{abstract}
\pacs{12.39.Fe,12.38.Aw,11.30.Rd,12.38.Lg,12.40.Ee}

\maketitle

\section{Introduction}

Thermodynamics of quantum chromodynamics (QCD) has been attracted a lot of attention during
the last three decades. Many interesting physical phenomena are related to it,
such as the equation of state of quark gluon plasma(QGP),
chiral symmetry breaking and restoration, deconfinement phase
transition, and so on. Thus, the study of the QCD thermodynamics becomes a basic problem in our understanding of the
strong interaction. Particularly, the deconfined QGP is expected to
be formed in ultrarelativistic heavy-ion collisions(HIC)
\cite{Shuryak2004,Gyulassy2005,Shuryak2005,Arsene2005,Back2005,Adams2005,Adcox2005,Blaizot2007}
and many experiments like those at the Relativistic Heavy Ion
Collision (RHIC) and at the Large Hadron Collider (LHC) have been built to
explore the nature of the QGP and to search for the critical point
of the phase transition, which enables many theoretical ideas
testable and makes the research area more exciting.

In this paper, we are going to study the thermodynamic properties of the effective chiral dynamical model(CDM)\cite{Dai:2003ip} of low energy QCD with spontaneous symmetry breaking via the dynamically generated effective composite Higgs potential of meson fields. The main assumption in such an effective CDM was based on an effective Lagrangian of chiral quarks with effective nonlinear-realized meson fields as the bosonized auxiliary fields, which may be resulted from
the Nambu-Jona-Lasinio(NJL) four quark interaction\cite{Nambu:1961tp} due to the strong interactions of
nonperturbative QCD at the low energy scale, then after considering the quantum loop contributions of quark fields by integrating
over the quark fields with the loop regularization (LORE) method\cite{wu1,Wu:2003dd}, the resulting effective chiral Lagrangian for the meson fields in the CDM\cite{Dai:2003ip}
has been turned out to provide a dynamically generated spontaneous symmetry breaking
mechanism for the $SU(3)_L\times SU(3)_R$ chiral symmetry. The key point for deriving the dynamically generated spontaneous symmetry breaking mechanism is the use of the LORE method which keeps the physically meaningful finite
quadratic term and meanwhile preserves the symmetries of
original theory. More specifically, the advantage of the LORE method is the introduction of two intrinsic energy scales without spoiling the basic symmetries of original theory, such intrinsic energy scales play the role of the characteristic energy
scale $M_c$ and the sliding energy scale $\mu_s$. Here
$M_c$ is the characterizing energy scale of nonperturbative QCD below which the effective quantum
field theory becomes meaningful to describe the low energy dynamics of
QCD, and $\mu_s$ reflects the low energy scale of QCD on
which the interesting physics processes are concerned. As a consequence,
it was shown that the resulting effective CDM of low energy QCD can lead to the consistent predictions for the light quark masses, quark
condensate, pseudoscalar meson masses and the lowest nonet scalar meson masses as well as their mixing at the leading order\cite{Dai:2003ip}.
Based on the success of the effective CDM at zero temperature, we will
show in this paper how the CDM can be extended to an effective chiral thermodynamic model (CTDM) at finite temperature by applying the closed-time-path Green function(CTPGF) approach\cite{Schwinger1961,Keldysh1965,Chou1985,Zhou1980,Rammer2007,Calzetta2008}, and how the CTDM enables us to describe the critical behavior of the low energy dynamics of QCD, and in particular to determine the critical temperature of the chiral symmetry breaking and restoration. This comes to our main motivation in the present paper. For our present purpose,  we will investigate the CTDM with considering only two flavors and ignoring the possible instanton effect which is known to account for the
the anomalous $U(1)_A$ symmetry breaking.

Following the almost same procedure in deriving the CDM for the composite meson fields, we will arrive at an
effective CTDM at finite temperature by adopting the CTPGF formalism. The key step
in the derivation is the replacement of the zero-temperature
propagator for the quark field with its finite temperature
counterpart in the CTPGF formalism which has been shown to be more suitable for
characterizing the nonequilibrium statistical processes\cite{Chou1985,Zhou1980,Rammer2007,Calzetta2008}.
We then obtain the dynamically generated effective composite Higgs potential at finite temperature,
its minimal condition leads to the gap equation at finite temperature and enables us to
explore the critical behavior and temperature for the chiral symmetry breaking and restoration. After fixing the free parameters in the
effective Lagrangian, we present our numerical predictions for the
temperature dependence of physically interesting quantities, such as
the vacuum expectation value $v_o(T)$, the pion decay constant $f_\pi(T)$, and
the masses of pseudoscalar and scalar mesons. The resulting critical
temperature is found to be $T_c \simeq 200$ MeV, which is consistent with the NJL model
prediction\cite{Klevansky1992,Hatsuda1994,Alkofer1996,Buballa2005}.

\section{Outline on Chiral Dynamical Model of Low Energy
QCD}\label{sec2}

Before exploring the thermodynamic behavior of the chiral dynamical
model, it is useful to have a brief review on its derivation of the effective chiral
Lagrangian at zero-temperature. For our present purpose with paying attention to the investigation of the
chiral symmetry restoration at finite temperature, we shall
consider the simple case with only two flavors of light quarks $u$ and $d$, and ignore the instanton
effects. For simplicity, we also assume the exact isospin symmetry for two light quarks $q=(u,d)^T$
by taking the equal mass $m_u=m_d=m$. Our discussion here is mainly following the previous paper by Dai and Wu\cite{Dai:2003ip}.

Let us begin with the QCD Langrangian for two light quarks
\begin{equation}\label{QCD}
{\cal L}_{QCD}=\bar{q}\gamma^\mu(i\partial_\mu+g_s G^a_\mu T^a)q -
\bar{q}M q-\frac{1}{2}tr G_{\mu\nu}G^{\mu\nu}
\end{equation}
where $q=(u,d)^T$ denotes the SU(2) doublet of two light quarks and the summation over
color degrees of freedom is understood. $G^a_\mu$ are the gluon
fields with SU(3) gauge symmetry and $g_s$ is the running coupling
constant. $M$ is the light quark mass matrix $M= diag(m_1,m_2)\equiv
diag(m_u,m_d)$. In the limit $m_i\to0$ ($i=1,2$), the Lagrangian has
the global $U(2)_L\times U(2)_R$ chiral symmetry. Due to the instanton effect,
the chiral symmetry $U(1)_L\times U(1)_R$ is broken down to the
diagonal $U(1)_V$ symmetry. This instanton effect can be expressed
by the effective interaction\cite{'tHooft1,'tHooft2}
\begin{equation}\label{instanton}
{\cal L}^{inst}=\kappa_{inst}e^{i\theta_{inst}}\det(-\bar{q}_R
q_L)+h.c.
\end{equation}
where $\kappa_{inst}$ is the constant containing the factor
$e^{-8\pi^2/g^2}$. Obviously, such an instanton term breaks the $U(1)_A$ chiral
symmetry. As our present consideration is paid to the phenomenon of the $SU(2)_L\times SU(2)_R$ chiral symmetry
breaking and its restoration at finite temperature, we will switch off the instanton effect
by simply setting $\kappa_{inst}=0$ in the following discussions.

The basic assumption of the CDM is that at the
chiral symmetry breaking scale ($\sim 1 GeV$) the effective
Lagrangian contains not only the quark fields but also the effective
meson fields describing bound states of strong interactions of
gluons and quarks. After integrating over the gluon fields at high
energy scales, the effective Lagrangian at low energy scale is
expected to have the following general form when keeping only the
lowest order nontrivial terms:
\begin{eqnarray}\label{Lag1}
{\cal L}_{eff}(q,\bar{q},\Phi)&= &\bar{q}\gamma^\mu i\partial_\mu
q+\bar{q}_L\gamma_\mu {\cal A}^\mu_L q_L+\bar{q}_R\gamma_\mu {\cal
A}_R^\mu q_R -[\bar{q}_L(\Phi-M)q_R+h.c.]\nonumber\\
&& +\mu_m^2 tr(\Phi M^\dagger+M\Phi^\dagger)-\mu^2_f tr
\Phi\Phi^\dagger
\end{eqnarray}
where $\Phi_{ij}$ are the effective meson fields which basically
correspond to the composite operators $\bar{q}_{Rj}q_{Li}$. ${\cal
A}_L$ and ${\cal A}_R$ are introduced as the external source fields.
It is noticed in Eq.(\ref{Lag1}) that the effective meson fields
$\Phi_{ij}$ are the auxiliary fields in the sense that there is no
kinetic term for them, which may explicitly be seen by integrating out $\Phi_{ij}$, we
then obtain the following effective Lagrangian of quarks:
\begin{eqnarray}
{\cal L}^{NJL}_{eff}(q,{\bar q}) &=& \bar{q}\gamma^\mu i\partial_\mu
q+\bar{q}_L\gamma_\mu {\cal A}^\mu_L q_L+\bar{q}_R\gamma_\mu {\cal
A}_R^\mu q_R\nonumber\\
&&- (\frac{\mu^2_m}{\mu_f^2}-1)({\bar q}_L M q_R +\bar{q}_R
M^\dagger q_L)+\frac{1}{\mu_f^2}{\bar q}_L q_R \bar{q}_R q_L
\end{eqnarray}
which arrives exactly at the Nambu-Jona-Lasinio(NJL)
model\cite{Nambu:1961tp} of effective four-quark interaction
with the quark mass matrix $(\frac{\mu_m^2}{\mu_f^2}-1)M$. When the
mass term is understood as the well-defined current quark mass term
in the QCD Lagrangian Eq.(\ref{QCD}), it is then clear that
$\mu_m^2/\mu_f^2 = 2$. Note that the high order terms with the
dimension above two for the effective meson fields are not included, which are
assumed to be small and generated in loop diagrams, so only the lowest order nontrivial fermionic interaction terms are
taken into account after integrating over the gluon field.

For our present purpose, we only focus on the scalar and pseudoscalar
mesons while the vector and axial vector sectors will not be considered here.
As the pseudoscalar mesons are known to be the would-be Goldstone
bosons, the effective chiral field theory is naturally to be
realized as a nonlinear model. Thus we may express the effective
meson fields $\Phi(x)$ into the following $2\times 2$ complex matrix
form:
\begin{eqnarray}\label{meson_def}
&& \Phi(x)\equiv \xi_L(x)\phi(x)\xi_R^\dagger(x),\qquad U(x)\equiv
\xi_L(x)\xi_R^\dagger(x)=\xi_L^2(x)=e^{i\frac{2\Pi(x)}{f}}\nonumber\\
&& \phi^\dagger(x)=\phi(x)=\sum^{3}_{a=0}\phi^a(x)T^{a},\qquad
\Pi^\dagger(x)=\Pi(x)=\sum^{3}_{a=0}\Pi^a(x)T^a
\end{eqnarray}
where $T^a$ ($a=0,1,2,3$) with $[T^a,T^b]=if^{abc}T^c$ and $2 tr T^a
T^b=\delta_{ab}$ are the four generators of $U(2)$ group. The fields
$\Pi^a(x)$ represent the pseudoscalar mesons and $\phi^a(x)$ the
corresponding scalar chiral partners. Where $f$ is known as the decay constant with mass
dimension.

The Lagrangian Eq.(\ref{Lag1}) with the definition of composite meson fields
$\Phi_{ij}(x)$ Eq.(\ref{meson_def}) is our starting point for the
derivation of the effective chiral Lagrangian for mesons. The
Lagrangian can be obtained by integrating over the quark fields and
the procedure for the derivation can be formally expressed in terms
of the generating functionals via the following relations
\begin{eqnarray}\label{formal}
\frac{1}{Z}\int [d G_\mu] [d q] [\bar{q}] e^{i\int d^4x {\cal
L}_{QCD}}&  = & \frac{1}{\bar Z} \int [d\Phi] [d q] [d\bar{q}] e^{i\int
d^4x {\cal L}_{eff}(q,{\bar q},\Phi)} \nonumber \\
& = & \frac{1}{Z_{eff}} \int[d\Phi]
e^{i\int d^4x {\cal L}_{eff}(\Phi)}
\end{eqnarray}

Let us first demonstrate the derivation of ${\cal L}_{eff}(\Phi)$.
In order to obtain the effective chiral Lagrangian for the meson fields,
we need to integrate over the quark fields (which is equivalent to calculate the Feynman diagrams of quark
loops) from the following chiral Lagrangian
\begin{equation}\label{Lag2}
{\cal L}^q_{eff}(q,{\bar q}) = \bar{q}\gamma^\mu i\partial_\mu
q+\bar{q}_L\gamma_\mu {\cal A}^\mu_L q_L+\bar{q}_R\gamma_\mu {\cal
A}_R^\mu q_R -[\bar{q}_L(\Phi-M)q_R+h.c.]+\chi\bar{q}q
\end{equation}
where we have introduced the real source field $\chi(x)$ for the
composite operator $\bar{q}q$ and the source term $\chi\bar{q}q$
which will be useful for the derivation of the chiral thermodynamic model,
while eventually the source field is taken to be zero $\chi =0$.

With the method of path integral, the effective Lagrangian of the
meson fields is evaluated by integrating over the quark fields
\begin{equation}
\int [d\Phi]exp\{i\int d^4x{\cal
L}^M\}=Z_0^{-1}\int[d\Phi][dq][d\bar{q}]exp\{i\int d^4x {\cal
L}^q_{eff}\}
\end{equation}
The functional integral of the right hand side is known as the determination of
the Dirac operator
\begin{equation}
\int[dq][d{\bar q}]exp\{i\int d^4 x {\cal L}^q_{eff}\}=\det(i{\cal
D}^\chi)
\end{equation}
To obtain the effective action, it is useful to go to Euclidean
space via the Wick rotation
\begin{equation}
\gamma_0\to i\gamma_4,\quad G_0\to i G_4,\quad x_0\to -i x_4
\end{equation}
and to define the Hermitian operator
\begin{eqnarray}
S^{M}_E &=& \int d^4x_E {\cal L}^M_E = \ln \det(i{\cal
D}^\chi_E)\nonumber\\
&=& \frac{1}{2}[\ln\det(i{\cal D}^\chi_E)+\ln\det(i{\cal
D}^\chi_E)^\dagger]+\frac{1}{2}[\ln\det(i{\cal
D}^\chi_E)-\ln\det(i{\cal
D}^\chi_E)^\dagger]\nonumber\\
&\equiv& S^{M}_{Re}+S^{M}_{Im}
\end{eqnarray}
with
\begin{eqnarray}
S^{M}_{E Re} &=& \int d^4x_E {\cal L}^M_{Re} =
\frac{1}{2}\ln\det(i{\cal D}^\chi_E(i{\cal D}^\chi_E)^\dagger)\equiv
\frac{1}{2}\ln\det \Delta^\chi_E-\ln Z_0\label{ChEF}\\
S^{ M}_{E Im} &=& \int d^4 x_E {\cal L}^M_{Im} = \frac{1}{2}
\ln\det(i{\cal D}^\chi_E/(i{\cal D}^\chi_E)^\dagger)\equiv
\frac{1}{2}\ln\det\Theta_E
\end{eqnarray}
where the imaginary part ${\cal L}^M_{Im}$ appears as a phase which
is related to the anomalous terms and will not be discussed in the
present paper. The operators in the Euclidean space are given by
\begin{eqnarray}
i{\cal D}^\chi_E &=& -i\gamma\cdot\partial-\gamma\cdot{\cal A}_L P_L
-\gamma\cdot{\cal A}_R P_R+\hat{\Phi}P_R+\hat{\Phi}^\dagger
P_L+\chi\nonumber\\
&=& i{\cal D}_E +\chi
\nonumber\\
(i{\cal D}^\chi_E)^\dagger &=& i\gamma\cdot\partial+\gamma\cdot{\cal
A}_R P_L +\gamma\cdot{\cal A}_L P_R+\hat{\Phi}^\dagger
P_R+\hat{\Phi}P_L+\chi\nonumber\\
&=& (i{\cal D}_E)^\dagger+\chi
\end{eqnarray}
with $\hat{\Phi}=\Phi-M$ and $P_{\pm}=(1\pm\gamma_5)/2$. $i{\cal
D}^\chi_E$, $(i{\cal D}^\chi_E)^\dagger$ and $\Delta^\chi_E$ are
regarded as matrices in coordinate space, internal symmetry space
and spin space. Noticing the following identity
\begin{equation}
\ln\det O= Tr\ln O
\end{equation}
with $Tr$ being understood as the trace defined via
\begin{equation}
Tr O = tr\int d^4 x <x|O|y>|_{x=y}
\end{equation}
Here $tr$ is the trace for the internal symmetry space and $<x|O|y>$
is the coordinate matrix element defined as
\begin{eqnarray}
<x|O_{ij}|y>=O^k_{ij}(x)\delta^4(x-y),\quad \delta^4(x-y) =
\int^{\infty}_{-\infty}\frac{d^4k}{(2\pi)^4}e^{ik\cdot(x-y)}
\end{eqnarray}
For the derivative operator, one has in the coordinate space
\begin{equation}
<x|\partial^\mu|y>=\delta^4(x-y) (-ik^\mu+\partial^\mu_y)
\end{equation}
With the above definitions, the operators $i{\cal D}^{k}_E$,
$(i{\cal D}^{k}_E)^\dagger$ and $\Delta^{k}_E$ in the
Euclidean space are given by
\begin{eqnarray}
i{\cal D}^{k}_E &=& -\gamma\cdot k
-i\gamma\cdot\partial-\gamma\cdot{\cal A}_L P_L -\gamma\cdot{\cal
A}_R P_R+\hat{\Phi}P_R+\hat{\Phi}^\dagger P_L+\chi\nonumber\\
&=& -\gamma\cdot k+i{\cal D}^\chi_E\\
(i{\cal D}_E^{ k})^\dagger &=& \gamma\cdot k+
i\gamma\cdot\partial+\gamma\cdot{\cal A}_R P_L +\gamma\cdot{\cal
A}_L P_R+\hat{\Phi}^\dagger P_R+\hat{\Phi}P_L+\chi\nonumber\\
&=& \gamma\cdot k+(i{\cal D}^\chi_E)^\dagger\\
\Delta_E^{k} &=&  k^2+\hat{\Phi}\hat{\Phi}^\dagger P_R
+\hat{\Phi}^\dagger\hat{\Phi}P_L-i\gamma\cdot D_E\Phi P_L
-i\gamma\cdot D_E \Phi^\dagger P_R\nonumber\\
&& -\sigma_{\mu\nu}{\cal F}_{R\mu\nu}P_L-\sigma_{\mu\nu}{\cal
F}_{L\mu\nu}P_R+(iD_{E\mu})(iD_{E\mu})+2k\cdot(iD_E)\nonumber\\
&& -i\gamma\cdot\partial\chi + \chi [i{\cal D}_E + (i{\cal D}_E)^\dagger ] +\chi^2\nonumber\\
&=& k^2+\Delta_E^\chi = k^2+\Delta_E+ \Delta_{\chi}
\end{eqnarray}
where
\begin{eqnarray}\label{partial}
iD_E\Phi &=& i\partial\Phi+{\cal A}_L\Phi-\Phi{\cal A}_R\\
iD_E &=& i\partial+{\cal A}_R P_L+{\cal A}_L P_R
\end{eqnarray}
and
\begin{eqnarray}
\Delta_E &\equiv& \hat{\Phi}\hat{\Phi}^\dagger P_R
+\hat{\Phi}^\dagger\hat{\Phi}P_L-i\gamma\cdot D_E\Phi P_L
-i\gamma\cdot D_E \Phi^\dagger P_R\nonumber\\
&& -\sigma_{\mu\nu}{\cal F}_{R\mu\nu}P_L-\sigma_{\mu\nu}{\cal
F}_{L\mu\nu}P_R+(iD_{E\mu})(iD_{E\mu})+2k\cdot(iD_E)\\
\Delta_{\chi} &\equiv& -i\gamma\cdot\partial\chi + \chi [i{\cal D}_E + (i{\cal D}_E)^\dagger ] +\chi^2 \\
\Delta_E^{\chi}&\equiv& \Delta_E+ \Delta_{\chi}
\end{eqnarray}

Thus the effective action is obtained as
\begin{equation}\label{ChEF2}
S^{M}_{E Re}=\frac{N_c}{2}\int d^4 x_E\int\frac{d^4 k}{(2\pi)^4} e^{ik(x-y)}
tr_{SF}\ln(k^2+\Delta^\chi_E)|_{x = y}-\ln Z_0
\end{equation}
where the subscripts $SF$ refer to the trace over the spin
and flavor indices and $N_c$ is the color number.

Before proceeding, we would like to mention some formulae which will be useful for the derivation of chiral thermodynamic model late on. By taking the functional derivative of Eq.(\ref{ChEF2}) with respect to the source field
$\chi$ at $\chi=0$, we have
\begin{eqnarray}\label{ChEFk}
\frac{\delta S^{M}_{E Re}}{\delta \chi(x)}|_{\chi =0} & = & \frac{N_c}{2}\int d^4 x_E
tr_{SF}\int\frac{d^4 k}{(2\pi)^4} e^{ik(x-y)} \left(-\gamma\cdot k + [i{\cal D}_E + (i{\cal D}_E)^\dagger ] \right)  \frac{1}{k^2+\Delta_E}|_{x = y} \\
& = &
 \frac{N_c}{2}\int d^4 x_E
tr_{SF}\int\frac{d^4 k}{(2\pi)^4} \left(-\gamma\cdot k + \gamma\cdot( {\cal A}_R -{\cal A}_L)\gamma_5 + \hat{\Phi} + \hat{\Phi}^{\dagger} \right)  \frac{1}{k^2+\Delta_E} \nonumber
\end{eqnarray}
from the right hand side one may pick up the quark propagator as $\chi$ is the source field for
the quark operator $\bar{q}(x)q(x)$. This can explicitly be shown by differentiating the effective
action with respect to $\chi$, which gives the coinciding limit of
the quark propagator $\lim_{x\to y}tr_{SF}\<T[q(x) \bar{q}(y)]\>$.

Alternatively, if taking the functional derivative of Eq.(\ref{ChEF2}) with respect to the operator
$\Delta_{\chi}$, we obtain another form of the formulae
\begin{equation}\label{ChEFd}
\frac{\delta S^{ M}_{E Re}}{\delta (\Delta_{\chi})_{ij}} =
\frac{N_c}{2}\int\frac{d^4k}{(2\pi)^4} \left(\frac{1} {k^2+\Delta_E+ \Delta_{\chi}}\right)_{ji}
\end{equation}
In other word, when functionally integrating over $\Delta_{\chi}$ and summing over the flavor and
spin degrees of freedom, we are led to the effective action
Eq.(\ref{ChEF2}). The physical action is yielded by taking the source field to be zero $\chi(x)=0$
\begin{equation}\label{ChEF3}
S^M_{E Re}=\frac{N_c}{2}\int d^4 x_E\int\frac{d^4 k}{(2\pi)^4}
tr_{SF}\ln(k^2+\Delta_E)-\ln Z_0
\end{equation}

To derive the effective chiral Lagrangian for the meson fields, we shall make the following redefinition for
$\Delta_E^k\equiv k^2+\Delta_E$
\begin{eqnarray}\label{separation}
\Delta_E^k&=&\Delta_0+\tilde{\Delta}_E \\
\Delta_0&\equiv&k^2+\bar{M}^2\\
\tilde{\Delta}_E&\equiv& (\hat{\Phi}\hat{\Phi}^\dagger-\bar{M}^2)P_R
+(\hat{\Phi}^\dagger\hat{\Phi}-\bar{M}^2)P_L -i\gamma\cdot D_E\Phi
P_L -i\gamma\cdot D_E\Phi^\dagger P_R\nonumber\\
&&-\sigma_{\mu\nu}{\cal F}_{R\mu\nu}P_L-\sigma_{\mu\nu}{\cal
F}_{L\mu\nu}P_R +(iD_{E\mu})(iD_{E\mu})+2k\cdot(iD_E)
\end{eqnarray}
where $\bar{M}$ is the supposed vacuum expectation values (VEVs) of
$\hat{\Phi}$, i.e.,
$<\hat{\Phi}>=\bar{M}=diag.(\bar{m}_u,\bar{m}_d)$ with
$\bar{m}_u=\bar{m}_d=\bar{m}$ under the exact isospin symmetry. Here
$\bar{m}_i=v_i-m_i$ is regarded as the dynamical quark masses, and
$v_i$ is supposed to be the VEVs of the scalar fields, i.e.,
$<\phi>=V=diag.(v_1,v_2)$ which will be determined from the minimal conditions
of the effective potential in the effective chiral Lagrangian
${\cal L}_{eff}(\Phi)$. With this convention, it is seen that the
minimal conditions of the effective potential are completely
determined by the lowest order terms up to the dimension four
$(\hat{\Phi}\hat{\Phi}^\dagger-\bar{M}^2)^2$ in the effective chiral
field theory of mesons.

By regarding $\tilde{\Delta}_E$ as the perturbative interaction term
and taking
$Z_0=(\det \Delta_0)^{\frac{1}{2}}$,
the effective action in the Euclidean space can be written as
\begin{eqnarray}
S^M_{ERe} &=& \frac{N_c}{2}\int d^4 x_E \int\frac{d^4 k}{(2\pi)^4}
tr_{SF} [\ln(\Delta_0+\tilde{\Delta}_E)-\ln\Delta_0]\nonumber\\
&=& \frac{N_c}{2}\int d^4 x_E
\int\frac{d^4k}{(2\pi)^4}tr_{SF}\ln(1+\frac{1}{\Delta_0}\tilde{\Delta}_E)\nonumber\\
&=& \frac{N_c}{2}\int d^4 x_E
\int\frac{d^4k}{(2\pi)^4}tr_{SF}\sum^\infty_{n=1}\frac{(-1)^{n+1}}{n}(\frac{1}{\Delta_0}\tilde{\Delta}_E)^n
\nonumber\\
&\simeq& \frac{N_c}{2}\int d^4 x_E \int\frac{d^4k}{(2\pi)^4}tr_{SF}
(\frac{1}{\Delta_0}\tilde{\Delta}_E-\frac{1}{2}\frac{1}{\Delta_0^2}\tilde{\Delta}_E^2)
\end{eqnarray}
Note that we only keep the first two terms in the expansion over
$\tilde{\Delta}_E$ since these are the only divergent terms in the
integration over the internal momentum $k$. Particularly, the
divergence degree of the first integral is quadratical while the
second logarithmic. In order to maintain the gauge invariance and
meanwhile keep the divergence behavior of the integral, we adopt the
loop regularization (LORE) method proposed in \cite{wu1,Wu:2003dd}
for the momentum integral
\begin{eqnarray}
I_2 &=& \int\frac{d^4k}{(2\pi)^4}\frac{1}{k^2+\bar{M}^2}\to
I^R_2=\frac{M_c^2}{16\pi^2} L_2(\frac{\mu^2}{M_c^2})\\
I_0 &=& \int\frac{d^4k}{(2\pi)^4}\frac{1}{(k^2+\bar{M}^2)^2}\to
I^R_0=\frac{1}{16\pi^2}L_0(\frac{\mu^2}{M_c^2})
\end{eqnarray}
with the consistent conditions for the tensor type divergent integrals
\begin{equation}
I^R_{2\mu\nu}=\frac{1}{2}g_{\mu\nu}I^R_2,\quad
I^R_{0\mu\nu}=\frac{1}{4}g_{\mu\nu}I^R_0
\end{equation}
where
\begin{eqnarray}
I_{2\mu\nu} = \int\frac{d^4k}{(2\pi)^4}\frac{k_{\mu}k_{\nu}}{(k^2+\bar{M}^2)^2},\qquad
I_{0\mu\nu} = \int\frac{d^4k}{(2\pi)^4}\frac{k_{\mu}k_{\nu}}{(k^2+\bar{M}^2)^3}
\end{eqnarray}
The two diagonal matrices $L_0=diag.(L^{(1)}_0,L^{(2)}_0)$ and
$L_2=diag.(L^{(1)}_2,L^{(2)}_2)$ are given by the following form:
\begin{eqnarray}\label{intg1}
L^{(i)}_0&=&\ln\frac{M_c^2}{\mu_i^2}-\gamma_\omega+y_0(\frac{\mu_i^2}{M_c^2})\\
L^{(i)}_2&=&1-\frac{\mu_i^2}{M_c^2}[\ln\frac{M_c^2}{\mu_i^2}-\gamma_\omega+1+y_2(\frac{\mu_i^2}{M_c^2})]
\end{eqnarray}
with
\begin{eqnarray}
y_0(x) &=& \int^x_0 d\sigma \frac{1-e^{-\sigma}}{\sigma},\quad
y_1(x) = \frac{1}{x} (e^{-x}-1+x),\quad
y_2(x) = y_0(x)-y_1(x)
\end{eqnarray}
Note that $M_c^2$ is the characteristic energy scale from which the
nonperturbative QCD effects start to play an important role and the
effective chiral field theory is considered to be valid below the
scale $M_c$. We have also introduced the definitions
\begin{equation}
\mu_i^2=\mu_s^2+\bar{m}_i^2, \quad \bar{m}_i=v_i-m_i
\end{equation}
with $\mu_s^2$ the sliding energy scale. It is usually taken to be at the energy scale at which the
physical processes take place, which is expected to be around the QCD scale $\Lambda_{QCD}$ for our present consideration.

With these analysis, the effective chiral Lagrangian can be
systematically obtained to be
\begin{eqnarray}\label{Lag_meson}
S^M_{ERe} &=& \frac{N_c}{16\pi^2}\int d^4 x_E tr_F
\{M_c^2L_2[(\hat{\Phi}\hat{\Phi}^\dagger-\bar{M}^2)+(\hat{\Phi}^\dagger\hat{\Phi}-\bar{M}^2)]\nonumber\\
&&-\frac{1}{2} L_0 [D_E\hat{\Phi}\cdot D_E\hat{\Phi}^\dagger+
D_E\hat{\Phi}^\dagger\cdot D_E\hat{\Phi}+
(\hat{\Phi}\hat{\Phi}^\dagger-\bar{M}^2)^2+(\hat{\Phi}^\dagger\hat{\Phi}-\bar{M}^2)^2]\}
\end{eqnarray}
where the trace over the spin indices give the factor
$2$ since our quark fields defined in Eq.(\ref{Lag1}) are Weyl
fermion fields, each of which has 2 degrees of freedom.

By transforming back to the Minkowski spacetime signature and adding
the extra terms in Eq.(\ref{Lag1}), we finally arrive at the
following effective chiral Lagrangian at zero temperature
\begin{eqnarray}
{\cal L}_{eff}(\Phi) &=& \frac{1}{2}\frac{N_c}{16\pi^2}tr_F
L_0[D_\mu\hat{\Phi}^\dagger D^\mu\hat{\Phi}
+D_\mu\hat{\Phi}D^\mu\hat{\Phi}^\dagger
-(\hat{\Phi}^\dagger\hat{\Phi}-\bar{M}^2)^2
-(\hat{\Phi}\hat{\Phi}^\dagger-\bar{M}^2)^2]\nonumber\\
&&+\frac{N_c}{16\pi^2}M_c^2 tr_F
L_2[(\hat{\Phi}^\dagger\hat{\Phi}-\bar{M}^2)
+(\hat{\Phi}\hat{\Phi}^\dagger-\bar{M}^2)]\nonumber\\
&&+\mu_m^2 tr_F(\Phi M^\dagger+M\Phi^\dagger)-\mu_f^2
tr\Phi\Phi^\dagger.
\end{eqnarray}

The derivation of the above effective chiral Lagrangian in an equivalent rotated basis\cite{Espriu:1989ff} is
given in Appendix \ref{appendix}, which may be more transparent for the spontaneous symmetry breaking with the composite Higgs-like scalar mesons.

\section{Derivation of Chiral Thermodynamic Model of
QCD}\label{title}

After a brief outline for the derivation of the effective chiral Lagrangian of the CDM
for mesons at zero temperature, it
is straightforward to incorporate the finite temperature effects
into the effective Lagrangian. The method for the derivation of finite
temperature effective Lagrangian is similar by applying for the closed-time-path
Green function(CTPGF) formalism. The CTPGF formalism, developed by
Schwinger~\cite{Schwinger1961} and Keldysh~\cite{Keldysh1965}, has
been used to solve lots of interesting problems in statistical
physics and condensed matter theory~\cite{Chou1985}. It is generally
believed that this technique is quite efficient in investigating the
nonequilibrium and finite temperature dynamical systems as this
formalism simultaneously incorporates both the statistical and dynamical
properties\cite{Chou1985,Zhou1980}. It has also
been used to treat a system of self-interacting bosons described by
$\lambda\phi^{4}$ scalar fields~\cite{Calzetta1988}. A brief introduction to this formalism is given in Appendix A. Readers who are not
familiar with the CTPGF formalism are refered to
the excellent review articles~\cite{Chou1985,Das:2000ft} and
monographs \cite{Calzetta2008,Rammer2007}.

As shown in the Appendix A, the main step for the derivation of
the effective action with finite temperature in the CTPGF formalism is to replace the field propagator with its finite temperature
counterpart\cite{Zhou1980,Chou1985,Quiros:1999jp,Das:2000ft}. Applying the CTPGF formalism to the propagator of the quark
fields in Eq.(\ref{ChEFk}), we arrive at the following result
\begin{eqnarray}
& & \frac{\delta S^{M}}{\delta \chi(x)}|_{\chi=0} =
\frac{N_c}{2}\int\frac{d^4k}{(2\pi)^4} tr_{SF} \left[-\gamma\cdot k + \left(i{\cal D}_E + (i{\cal D}_E)^\dagger \right) \right] \nonumber \\
& & \left [\left(
\begin{array}{cc}
\frac{1}{k^2+\Delta_E} & 2\pi i\theta(-k_4)\delta(k^2+\Delta_E) \\
2\pi i\theta(k_4)\delta(k^2+\Delta_E) & -\frac{1}{k^2+\Delta_E} \\
\end{array}
\right) - 2\pi i n_F(\omega)\delta(k^2+\Delta_E) \left(
\begin{array}{cc}
1 & 1\\
1 & 1
\end{array}\right)\right]
\end{eqnarray}
where $\omega$ is defined as the effective energy
$\omega\equiv\sqrt{\vec{k}^2+\Delta_E}$ and $n_F(\omega)$
represents the Fermi-Dirac distribution function
\begin{equation}
n_F(\omega)\equiv \frac{1}{e^{\beta\omega}+1}
\end{equation}
For our present purpose, we only need to calculate the first component of the
effective action since it is the only one which is related to the
causal propagation\cite{Quiros:1999jp,Das:2000ft}. For convenience, we may adopt the following alternative formula which is similar to Eq.(\ref{ChEFd}) at zero temperature
\begin{eqnarray}
\frac{\delta S^{ M}_{E Re}}{\delta \Delta_{\chi}}&=&\frac{N_c}{2}\int\frac{d^4
k}{(2\pi)^4} [\frac{1}{k^2+\Delta_E+ \Delta_{\chi} }-2\pi i
n_F(\omega)\delta(k^2+\Delta_E+ \Delta_{\chi} )]\nonumber\\
&=& \frac{N_c}{2} [\int\frac{d^4 k}{(2\pi)^4}
\frac{1}{k^2+\Delta_E+ \Delta_{\chi} } -\int\frac{d^3
k}{(2\pi)^3}n_F(\omega)\frac{1}{\sqrt{\vec{k}^2+\Delta_E+ \Delta_{\chi} }} ]
\end{eqnarray}
By functionally integrating over $\Delta_{\chi}$ and summing over the spin and
flavor indices, we then obtain the effective action for the CTDM
\begin{eqnarray}
S^M_{E Re} &=&  \frac{N_c}{2}\int d^4 x_E tr_{SF}[ \int\frac{d^4
k}{(2\pi)^4}\ln(k^2+\Delta_E)+\frac{1}{\beta}\int \frac{d^3
k}{(2\pi)^3}\ln(1+e^{-\beta\sqrt{\vec{k}+\Delta_E}})] \nonumber\\
&&-\ln Z_0
\end{eqnarray}
where we have put the source field $\chi(x)=0$ in the end  of the
calculation.

Separating $\Delta_E^k$ as in Eq.(\ref{separation}) and
identifying $\tilde{\Delta}_E$ as the perturbation part, we then make the
expansion in terms of $\tilde{\Delta}_E$. Here we only keep the lowest order terms
as they are the only ones relevant to our present
discussion. Also, we take
\begin{equation}
\ln Z_0= \frac{N_c}{2}\int d^4 x_E tr_{SF}[ \int\frac{d^4
k}{(2\pi)^4}\ln(k^2+\bar{M}^2)+\frac{1}{\beta}\int \frac{d^3
k}{(2\pi)^3}\ln(1+e^{-\beta\sqrt{\vec{k}+\bar{M}^2}})]
\end{equation}
which provides the cancelation for the infinite zero-point energy.

Thus, the effective action with leading terms in the Euclidean space can be written as
\begin{eqnarray}\label{ChEF8}
&S^M_{E Re}& \simeq \frac{N_c}{2}\int d^4 x_E
tr_{SF}\{[\int\frac{d^4
k}{(2\pi)^4}\frac{1}{k^2+\bar{M}^2}-\int\frac{d^3
k}{(2\pi)^3}\frac{1}{\sqrt{\vec{k}^2+\bar{M}^2}(e^{\beta\sqrt{\vec{k}^2+\bar{M}^2}}+1)}]
\tilde{\Delta}_E\nonumber\\
&&- \frac{1}{2}[\int\frac{d^4k}{(2\pi)^4}\frac{1}{(k^2+\bar{M}^2)^2}
-\int\frac{d^3k}{(2\pi)^3}\frac{\beta\sqrt{\vec{k}^2+\bar{M}^2}
e^{\beta\sqrt{\vec{k}^2+\bar{M}^2}}+e^{\beta\sqrt{\vec{k}^2+\bar{M}^2}}+1}{2(\vec{k}^2+\bar{M}^2)^{3/2}
(e^{\beta\sqrt{\vec{k}^2+\bar{M}^2}}+1)^2}]\tilde{\Delta}^2_E\}\nonumber\\
&&= \frac{N_c}{16\pi^2} \int d^4 x_E
 tr_F\{M_c^2 L_2(T)[(\hat{\Phi}\hat{\Phi}^\dagger-\bar{M}^2)+(\hat{\Phi}^\dagger\hat{\Phi}-\bar{M}^2)]\nonumber\\
&&-\frac{1}{2} L_0(T) [D_E\hat{\Phi}\cdot D_E\hat{\Phi}^\dagger+
D_E\hat{\Phi}^\dagger\cdot D_E\hat{\Phi}+
(\hat{\Phi}\hat{\Phi}^\dagger-\bar{M}^2)^2+(\hat{\Phi}^\dagger\hat{\Phi}-\bar{M}^2)^2]\}
\end{eqnarray}
where we have defined
\begin{eqnarray}\label{tt}
L_0(T) &\equiv& L_0(\frac{\mu^2}{M_c^2})-\frac{1}{\pi}\int d^3 k
\frac{\beta\sqrt{\vec{k}^2+\bar{M}^2}
e^{\beta\sqrt{\vec{k}^2+\bar{M}^2}}+e^{\beta\sqrt{\vec{k}^2+\bar{M}^2}}+1}{(\vec{k}^2+\bar{M}^2)^{3/2}
(e^{\beta\sqrt{\vec{k}^2+\bar{M}^2}}+1)^2} \nonumber \\
L_2(T) &\equiv&
L_2(\frac{\mu^2}{M_c^2})-\frac{4}{\pi M_c^2}\int d^3 k
\frac{1}{\sqrt{\vec{k}^2+\bar{M}^2}(e^{\beta\sqrt{\vec{k}^2+\bar{M}^2}}+1)}
\end{eqnarray}
and used the results
\begin{eqnarray}
tr_{S}\tilde{\Delta}_E &=& 2
[(\hat{\Phi}\hat{\Phi}^\dagger-\bar{M}^2)+(\hat{\Phi}^\dagger\hat{\Phi}-\bar{M}^2)]\\
tr_{S}\tilde{\Delta}_E^2 &=& 2[D_E\hat{\Phi}\cdot
D_E\hat{\Phi}^\dagger+ D_E\hat{\Phi}^\dagger\cdot D_E\hat{\Phi}+
(\hat{\Phi}\hat{\Phi}^\dagger-\bar{M}^2)^2+(\hat{\Phi}^\dagger\hat{\Phi}-\bar{M}^2)^2]
\end{eqnarray}

Finally, transforming the action to the Minkowski spacetime and adding
the extra terms in Eq.(\ref{Lag1}), we arrive at the following effective chiral
Lagrangian at finite temperature for the composite meson fields
\begin{eqnarray}
{\cal L}_{eff}(\Phi) &=& \frac{1}{2}\frac{N_c}{16\pi^2}tr_F
L_0(T) [D_\mu\hat{\Phi}^\dagger D^\mu\hat{\Phi}
+D_\mu\hat{\Phi}D^\mu\hat{\Phi}^\dagger
-(\hat{\Phi}^\dagger\hat{\Phi}-\bar{M}^2)^2
-(\hat{\Phi}\hat{\Phi}^\dagger-\bar{M}^2)^2]\nonumber\\
&&+\frac{N_c}{16\pi^2}M_c^2 tr_F
L_2(T) [(\hat{\Phi}^\dagger\hat{\Phi}-\bar{M}^2)
+(\hat{\Phi}\hat{\Phi}^\dagger-\bar{M}^2)]\nonumber\\
&&+\mu_m^2(T) tr_F(\Phi M^\dagger+M\Phi^\dagger)-\mu_f^2(T)
tr\Phi\Phi^\dagger
\end{eqnarray}
where $L_0(T) $ and $L_2(T) $ are given in Eq.(\ref{tt}). Note that the initial mass scale $\mu_f$ ($\mu_m$) characterizes the nonperturbative gluon effect at zero temperature. At the finite temperature, it is expected that the mass scale $\mu_f$($\mu_m$) is in general temperature dependent, which will be seen more clear below.

\section{Dynamical Symmetry Breaking and Thermodynamic Properties of CTDM}

Let us now focus on the dynamically generated effective composite Higgs potential of meson fields, which can be reexpressed
as the following general form
\begin{eqnarray}
V_{eff}(\Phi) &=& -tr_F \hat{\mu}_m^2(T)(\Phi M^\dagger+ M
\Phi^\dagger)+\frac{1}{2}tr_F\hat{\mu}^2_f(T)
(\Phi\Phi^\dagger+\Phi^\dagger \Phi)\nonumber\\
&&+\frac{1}{2}tr_F\lambda(T)[(\hat{\Phi}\hat{\Phi}^\dagger)^2+(\hat{\Phi}^\dagger\hat{\Phi})^2]
\end{eqnarray}
with $\hat{\mu}^2_f(T)$, $\hat{\mu}_m^2(T)$ and $\lambda(T)$ the
three diagonal matrices
\begin{eqnarray}
\hat{\mu}_f^2(T) &\equiv&
\mu_f^2(T)-\frac{N_c}{8\pi^2}(M_c^2L_2(T)+\bar{M}^2L_0(T))\\
\hat{\mu}_m^2(T) &\equiv&
\mu_m^2(T)-\frac{N_c}{8\pi^2}(M_c^2L_2(T)+\bar{M}^2L_0(T))\\
\lambda(T) &\equiv& \frac{N_c}{16\pi^2}L_0(T)
\end{eqnarray}
Taking the nonlinear realization
$\Phi(x)=\xi_L(x)\phi(x)\xi_R^\dagger(x)$ with supposing that the
minimal of the above effective potential occurs at the point
$<\phi>=V(T)=diag.(v_1(T),v_2(T))$, we can write the scalar fields as follows
\begin{equation}
\phi(x)=V(T)+\varphi(x)
\end{equation}
where the VEVs may be written in terms of the following general form:
\begin{equation}
v_i(T)=v_o(T)+\beta_o m_i\quad\quad i=1,2 \quad or \quad i=u,d
\end{equation}
For the equal mass $m_u=m_d=m$ considered in our present case, it leads to
the general VEVs $v_1(T)=v_2(T)=v(T)$ and the single form $v(T)=v_o(T)+\beta_o
m$.

By differentiating the effective composite Higgs potential of the scalar meson field at the VEV $v(T)$, we then obtain
the minimal conditions:
\begin{equation}\label{mini cond}
-\hat{\mu}^2_f(T)_i v(T)_i+ \hat{\mu}^2_m(T)_i m_i-2\lambda(T)_i
\bar{m}^3(T)_i=0
\end{equation}
with equal mass of two flavor quarks, it reduces to one minimal condition. For convenience of discussions, it is
helpful to decompose $\mu^2(T)$, $\hat{\mu}^2_f(T)$,
$\hat{\mu}_m^2(T)$ and $\lambda(T)$ into two parts with one part independent of the current quark mass $m$.
Practically, it can be done by making an expansion with respect to
the current quark masses
\begin{eqnarray}
&& \mu^2(T)=\mu_o^2(T)+2(\beta_o-1)v_o(T) \tilde{m},\quad
\mu_o^2(T)=\mu_s^2+v_o^2(T),\nonumber\\
&&
\tilde{m}(T)=m[1+(\beta_o-1)m/(2v_o(T))]\\
&& \hat{\mu}_f^2(T) =
\bar{\mu}_f^2(T)+2\mu_{fo}(T)\tilde{m}(T)[1+\sum_{k=1}\alpha_k(T)(\frac{\tilde{m}(T)}{\mu_o(T)})^k(\beta_o-1)^k]\\
&& \hat{\mu}_m^2(T) =
\bar{\mu}_m^2(T)+2\mu_{fo}(T)\tilde{m}(T)[1+\sum_{k=1}\alpha_k(T)(\frac{\tilde{m}(T)}{\mu_o(T)})^k(\beta_o-1)^k]\\
&& \lambda(T)=\bar{\lambda}(T)-\lambda_o
\sum_{k=1}\beta_k(T)(\frac{\tilde{m}(T)}{\mu_o(T)})^k(\beta_o-1)^k,
\quad \lambda_o=\frac{N_c}{16\pi^2}
\end{eqnarray}

By keeping only the nonzero leading terms in the expansion of
current quark masses, we then obtain the following constraints from
the minimal condition Eq.(\ref{mini cond})
\begin{equation}\label{mini1}
\bar{\mu}_f^2(T)+2\bar{\lambda}(T)v_o^2(T)=0
\end{equation}
Here the temperature-dependent parameters $\bar{\mu}_m^2(T)$,
$\bar{\mu}_f^2(T)$ and $\bar{\lambda}(T)$ are related to the initial
parameters in the effective potential and the characteristic energy
scale via the following relations
\begin{eqnarray}
\bar{\mu}_f^2(T) &=&
\mu_f^2(T)-\frac{N_c}{8\pi^2}(M_c^2\bar{L}_2(T)+v_o^2(T)\bar{L}_0(T))\\
\bar{\mu}_m^2(T) &=&
\mu_m^2(T)-\frac{N_c}{8\pi^2}(M_c^2\bar{L}_2(T)+v_o^2(T)\bar{L}_0(T))\\
\bar{\lambda}(T) &=& \frac{N_c}{16\pi^2}\bar{L}_0(T)
\end{eqnarray}
where $\bar{L}_0(T)$ and $\bar{L}_2(T)$ represent the leading order
expansion of $L_0(T)$ and $L_2(T)$ with respect to $m$. Explicitly,
they are given by
\begin{eqnarray}
\bar{L}_0(T) &\equiv&
L_0(\frac{\mu_o^2(T)}{M_c^2})-\frac{1}{\pi}\int d^3 k
\frac{\beta\sqrt{\vec{k}^2+v_o^2(T)}
e^{\beta\sqrt{\vec{k}^2+v_o^2(T)}}+e^{\beta\sqrt{\vec{k}^2+v_o^2(T)}}+1}{(\vec{k}^2+v_o^2(T))^{3/2}
(e^{\beta\sqrt{\vec{k}^2+v_o^2(T)}}+1)^2}\\
\bar{L}_2(T)&\equiv& L_2(\frac{\mu_o^2(T)}{M_c^2})-\frac{4}{\pi
M_c^2}\int d^3 k
\frac{1}{\sqrt{\vec{k}^2+v_o^2(T)}(e^{\beta\sqrt{\vec{k}^2+v_o^2(T)}}+1)}
\end{eqnarray}
With these definitions of parameters, the minimal condition
Eq.(\ref{mini1}) is transformed into the following form
\begin{eqnarray}\label{gap eqn}
\mu_f^2(T) &=& \frac{N_c}{8\pi^2}M_c^2 \bar{L}_2(T)\nonumber\\
&=&
\frac{N_c}{8\pi^2}[M_c^2-\mu_o^2(T)(\ln\frac{M_c^2}{\mu_o^2(T)}-\gamma_\omega+1+y_2(\frac{\mu_o^2(T)}{M_c^2}))]\nonumber\\
&& -\frac{2N_c}{\pi^2}\int^\infty_0
dk\frac{k^2}{\sqrt{k^2+v_o^2(T)}(e^{\beta\sqrt{k^2+v_o^2(T)}}+1)}
\end{eqnarray}
which is the gap equation at finite temperature. In order to obtain the
critical temperature, let us make a simple assumption for the
temperature dependence of the mass scale $\mu_f^2(T)$
\begin{equation}\label{mu_f}
\mu_f^2(T) = \gamma_o v_o^2(T)
\end{equation}
with $\gamma_o$ a temperature independent constant. The reason for this assumption will become manifest below from the thermodynamic property of the pion meson mass. In fact, recall that the appearance of $\mu_f^2$ in the effective
Lagrangian Eq.(\ref{Lag1}) can be traced back to the integration
over the gluon fields. In principle, it can be calculated from QCD
and given in terms of the QCD parameters $g_s(\mu)$ and
$\Lambda_{QCD}$. So the thermodynamic property
of $\mu_f^2$ is expected to be obtained from the detailed analysis of gluon dynamics at finite temperature.
The above simple assumption means that the gluon thermodynamics has the same temperature dependence as the chiral thermodynamics of quark
condensate.

With such an assumption and based on the chiral thermodynamic gap equation, we are able to calculate the
critical temperature for the chiral symmetry restoration. Suppose
that at the critical temperature the VEV $v_o(T)$ approaches to vanish,
so does the $\mu_f^2(T)$, then the gap equation becomes
\begin{eqnarray}
0 &=&
\frac{N_c}{8\pi^2}[M_c^2-\mu_s^2(\ln\frac{M_c^2}{\mu_s^2}-\gamma_\omega+1+y_2(\frac{\mu_s^2}{M_c^2})]-\frac{2N_c}{\pi^2}\int^\infty_0
d
k \frac{k}{e^{\beta k}+1}\nonumber\\
&=&
\frac{N_c}{8\pi^2}[M_c^2-\mu_s^2(\ln\frac{M_c^2}{\mu_s^2}-\gamma_\omega+1+y_2(\frac{\mu_s^2}{M_c^2})]
-\frac{2N_c}{\pi^2}T^2\int^\infty_0 d
k^\prime \frac{k^\prime}{e^{k^\prime}+1}\nonumber\\
&=&\frac{N_c}{8\pi^2}[M_c^2-\mu_s^2(\ln\frac{M_c^2}{\mu_s^2}-\gamma_\omega+1+y_2(\frac{\mu_s^2}{M_c^2})]
-\frac{2N_c}{12}T_c^2
\end{eqnarray}
where $k^\prime$ is defined as $k^\prime=\beta k$ and we have used
the result
\begin{equation}
\int^\infty_0 d k^\prime
\frac{k^\prime}{e^{k^\prime}+1}=\frac{\pi^2}{12}
\end{equation}
Thus, the critical temperature for the chiral symmetry restoration
is given by
\begin{equation}\label{Tc}
T_c =
\sqrt{\frac{3}{4\pi^2}[M_c^2-\mu_s^2(\ln\frac{M_c^2}{\mu_s^2}-\gamma_\omega+1+y_2(\frac{\mu_s^2}{M_c^2}))]}
\end{equation}
which shows that the critical temperature is characterized by the quadratic term evaluated in the LORE method, which differs from the dimensional regularization where the quadratic term is in general suppressed.

So far, we have explicitly shown the mechanism of dynamical
spontaneous chiral symmetry breaking and its restoration at finite temperature.

We are now going to present the explicit expressions for the masses of the scalar
mesons, pseudoscalar mesons and/or light quarks. To be
manifest, let us first write down the scalar and pseudoscalar meson
matrices
\begin{eqnarray}
\sqrt{2}\varphi= \left(
                   \begin{array}{cc}
                     \frac{a_0^0}{\sqrt{2}}+\frac{\sigma}{\sqrt{2}} & a_0^+ \\
                     a_0^- & -\frac{a_0^0}{\sqrt{2}}+\frac{\sigma}{\sqrt{2}}  \\
                   \end{array}
                 \right)
\end{eqnarray}
and
\begin{eqnarray}
\sqrt{2}\Pi= \left(
                   \begin{array}{cc}
                     \frac{\pi^0}{\sqrt{2}}+\frac{\eta'}{\sqrt{2}} & \pi^+ \\
                     \pi^- & -\frac{\pi^0}{\sqrt{2}}+\frac{\eta'}{\sqrt{2}}  \\
                   \end{array}
                 \right)
\end{eqnarray}

Keeping to the leading order of the current quark masses, we have
\begin{eqnarray}\label{pscalar}
m_{\pi^{0,\pm}}^2(T)=m_{\eta'}^2(T)\simeq
\frac{2\mu_P^3(T)}{f_\pi^2(T)}(m_u+m_d)=\frac{4\mu_P^3(T)}{f_\pi^2(T)}m
\end{eqnarray}
for the pseudoscalar mesons, and
\begin{equation}\label{scalar}
m_{a_0^{0,\pm}}^2(T) = m_{\sigma}^2(T) \simeq
3(\bar{m}_u^2(T)+\bar{m}_d^2(T))=6\bar{m}^2(T)
\end{equation}
for the scalar mesons. Where $\mu_P^3$ is given by
\begin{equation}\label{mu_P}
\mu_P^3(T)=(\bar{\mu}^2_m(T)+2\bar{\lambda}(T)v_o^2(T))v_o(T)=\mu_f^2(T)
v_o(T)=\gamma_o v_o^3(T)
\end{equation}
where we have used the minimal condition Eq.(\ref{mini1}) and the
relation Eq.(\ref{mu_f}).

Note that in obtaining the above results for the scalar and pseudoscalar meson masses, the
SU(2) triplet and singlet mesons have the common masses: $m_{a_0}^2=m_{\sigma}^2$ and
$m^2_{\pi}=m^2_{\eta'}$, which is the reflection of the present assumption of the
exact $U(1)_{A}$ symmetry. However, as we discussed in Sec. 2, in
the real world such a symmetry gets quantum anomalous and will be broken down by
the instanton effects, which is ignored in our present consideration.

\section{Predictions with Input Parameters at Low Energy and Critical Temperature in CTDM}

In order to make numerical predictions for the temperature dependence of the masses of the
light scalar and pseudoscalar mesons, it needs to fix the values of input parameters in
the effective chiral Lagrangian with finite temperature. There are in general five parameters: $\mu_f^2$ ($\mu_m^2$), $M_c^2$, $\mu_s^2$, $v_o$, and a universal current quark mass $m$. To fix the parameters, we shall use the constraints at low energy with zero temperature.

In general, the minimal condition Eq.(\ref{mini cond}) with different quark masses will lead to two
constraints by expanding the equation with respect to the current
quark mass up to the order of $m^2$. For the equal mass case, we get the following minimal condition
\begin{eqnarray}
&& \bar{\mu}_f^2+2\bar{\lambda}v_o^2=0\label{mini cond1}
\end{eqnarray}
with
\begin{eqnarray}
& &
\frac{\lambda_o}{\bar{\lambda}}[(\frac{2v_o^2}{\mu_o^2}-1)(1-\frac{v_o^2}{3\mu_o^2})
-\frac{2v_o}{3\mu_o}\alpha_1(1-r)+r]=1\label{mini cond2} \\
&& r\equiv\frac{\mu_s^2}{\mu_o^2}-\frac{\mu_o^2}{M_c^2}[1-\frac{\mu_s^2}{\mu_o^2}+O(\frac{\mu_o^2}{M_c^2})]\\
&& \alpha_1(1-r) \equiv
\frac{2v_o}{\mu_o}[\frac{\mu_s^2}{2\mu_o^2}+O(\frac{\mu_o^2}{M_c^2})]
\end{eqnarray}
Note that in obtaining the above result one needs to keep to
the order of $m^2$ in the current quark mass expansion.

As we have shown in Sec.II that in order to have well-defined QCD
current quark masses, it requires that
\begin{equation}\label{mm}
(\frac{\mu_m^2}{\mu_f^2}-1)M=M,\quad i.e.\quad \mu_m^2=2\mu_f^2
\end{equation}
which fixes the parameter
\begin{equation}\label{beta}
\beta_o=\frac{\mu_m^2}{\mu_f^2}=2
\end{equation}

Also, from the original Lagrangian of chiral dynamical model
Eq.(\ref{Lag1}), the auxiliary fields $\Phi_{ij}$ are found to be
given by the quark fields as follows
\begin{equation}
\Phi_{ij}=-\frac{1}{\mu^2_f}\bar{q}_{Rj}q_{Li}+\frac{\mu_m^2}{\mu^2_f}M_{ij}
\end{equation}
By assuming that the quark condensation is almost flavor independent,
i.e., $<\bar{u}u>\simeq<\bar{d}d>$, and combining the condition Eq.(\ref{mm}), then the dynamical quark
masses take the simple form
\begin{equation}
\bar{m}=v-m=v_o+(\beta_o-1)m=v_o+m
\end{equation}
which may be identified with the expected constituent quark masses
after dynamically spontaneous symmetry breaking, and $v_o$ is caused
by the quark condensation
\begin{equation}\label{cond}
v_o=-\frac{1}{2\mu_f^2}\<\bar{q}q\>,\quad q=u,d
\end{equation}

To determine the remaining parameters, we consider the following
constraints. There are two constraints arising from the pseudoscalar
sector. One is from the normalization of the kinetic terms.
\begin{equation}\label{pi decay}
\bar{\lambda}v_o^2={f_\pi^2\over 4}
\end{equation}
After some manipulation, the equation can be transformed into the
following form
\begin{equation}\label{ps kine}
L_0(\frac{\mu_o^2}{M_c^2})v_o^2=\frac{(4\pi f_\pi)^2}{4N_c}\equiv
\bar{\Lambda}_f^2\simeq (340 MeV)^2
\end{equation}
where we have used the pion decay constant $f_\pi\simeq 94$MeV.

The other comes from the current quark mass and pion mass via the relation Eq.(\ref{pscalar}). Taking the pion mass
$m_{\pi^{0,\pm}}\simeq 139$ MeV and the VEV $v_o = 350^{+20}_{-20}$ MeV or alternatively the current quark mass
$m=4.76_{+0.08}^{-0.04}$MeV as the inputs, we obtain
the following relation
\begin{eqnarray}\label{condensation}
2\mu_f^2 v_o =-\<\bar{q}q\> =\frac{m_{\pi^{0,\pm}}^2
f_{\pi}^2}{2m}=(262^{+1}_{-2} MeV)^3
\end{eqnarray}

With the above relations and constraints Eqs.(\ref{mini cond1}), (\ref{mini
cond2}), (\ref{mm}), (\ref{cond}),(\ref{ps kine}),
(\ref{condensation}), all the parameters can be completely determined
\begin{eqnarray}\label{det}
&& v_o\simeq 350^{+20}_{-20}MeV\nonumber\\
&& M_c\simeq 881_{+57}^{-32}MeV,\quad \mu_s\simeq 312_{+11}^{-3}MeV\nonumber\\
&& \mu_m^2=2\mu_f^2=(226_{+7}^{-5}MeV)^2\nonumber\\
&& \beta_o=2, \quad \gamma_o={\mu_f^2\over v_o^2}=0.209_{+0.041}^{-0.031}\nonumber\\
&& \<\bar{q}q\> = -(262^{+1}_{-2} MeV)^3
\end{eqnarray}

With these parameters, we can immediately obtain the critical
temperature for the chiral symmetry restoration
\begin{equation}\label{Tc}
T_c =
\sqrt{\frac{6}{8\pi^2}[M_c^2-\mu_s^2(\ln\frac{M_c^2}{\mu_s^2}-\gamma_\omega+1+y_2(\frac{\mu_s^2}{M_c^2}))]}\simeq
200_{+15}^{-9}MeV
\end{equation}
which is consistent with NJL model prediction
\cite{Hatsuda1994,Klevansky1992,Alkofer1996,Buballa2005}.

\section{ Chiral Symmetry Restoration and Critical Phase Transition of Low Energy QCD }

In this section, we will present numerical predictions based on the CTDM. Especially we will show the
thermodynamic behavior of the VEV $v_o(T)$, pion decay constant
$f_\pi(T)$, the quark condensate $\<\bar{q}q\>(T)$ and the masses of
pseudoscalar mesons $m_{\pi^{0,\pm}}(T)$. Their temperature dependence and the properties of critical phase transition are plotted in all the
diagrams with adopting the central values of the
quantities listed in Eq.(\ref{det}).

\begin{figure}[ht]
\begin{center}
  \includegraphics{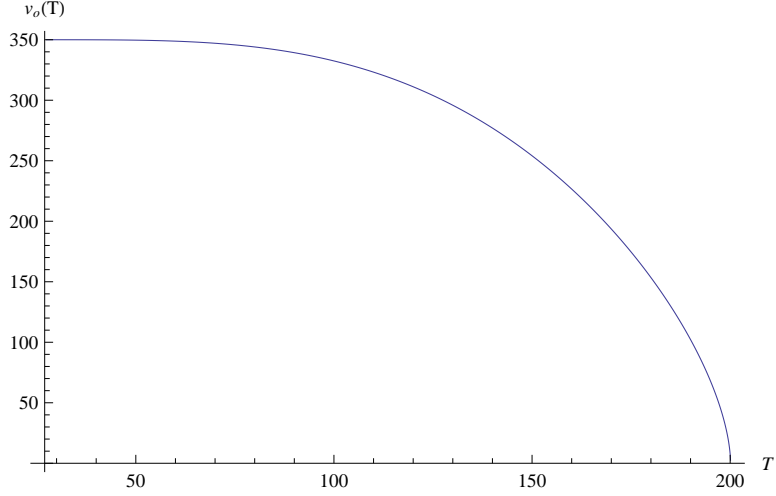}\\
  \caption{Temperature dependence of VEV $v_o$}\label{v01}
\end{center}
\end{figure}

From the gap equation Eq.(\ref{gap eqn}), we can numerically solve
the vacuum expectation value $v_o(T)$ at any finite temperature until the
critical temperature where $v_o(T)$ approaches to vanish. The result is
shown in Fig.(\ref{v01}).

\begin{figure}[ht]
\begin{center}
  \includegraphics{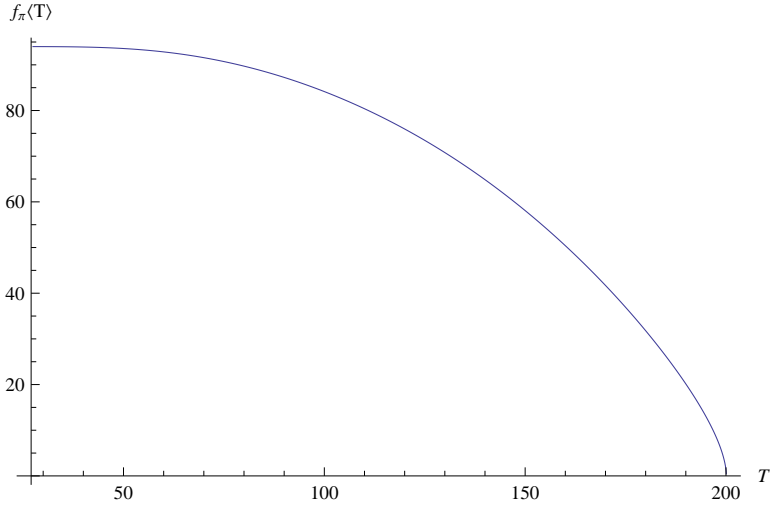}\\
  \caption{Temperature dependence of the pion decay constant}\label{fpi1}
\end{center}
\end{figure}

By the normalization of kinetic terms of pseudoscalar sector
Eq.(\ref{pi decay}), we can obtain the expression determining the
pion decay constant at finite temperature
\begin{equation}
f_\pi(T) =
\sqrt{4\bar{\lambda}(T)v_o^2(T)}=2v_o(T)\sqrt{\frac{N_c}{16\pi^2}\bar{L}_0(T)}
\end{equation}
which is presented in Fig.(\ref{fpi1})

\begin{figure}[ht]
\begin{center}
  \includegraphics{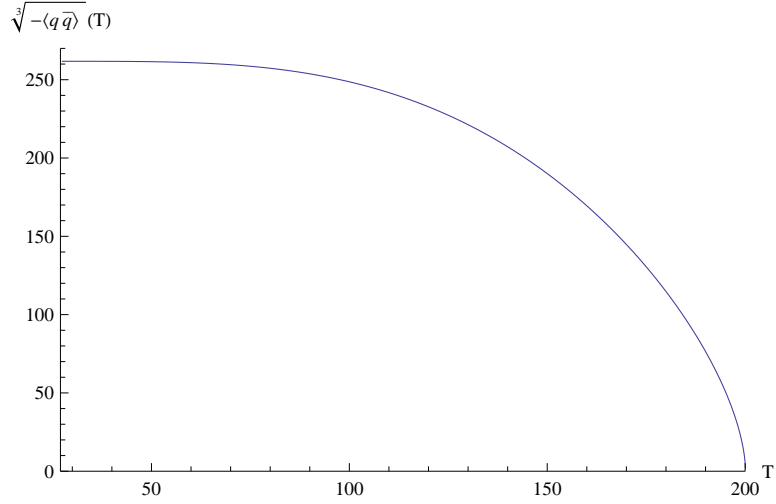}\\
  \caption{Temperature dependence of the quark condensate}\label{cond2}
\end{center}
\end{figure}

Furthermore, the quark condensate $\<\bar{q}q\>(T)$ is given by:
\begin{equation}
\<\bar{q}q\>(T) =-2\mu_f^2(T) v_o(T)=-2\gamma_o v_o^3(T)
\end{equation}
its variation with respect to temperature is displayed in
Fig.(\ref{cond2})

\begin{figure}[ht]
\begin{center}
  \includegraphics{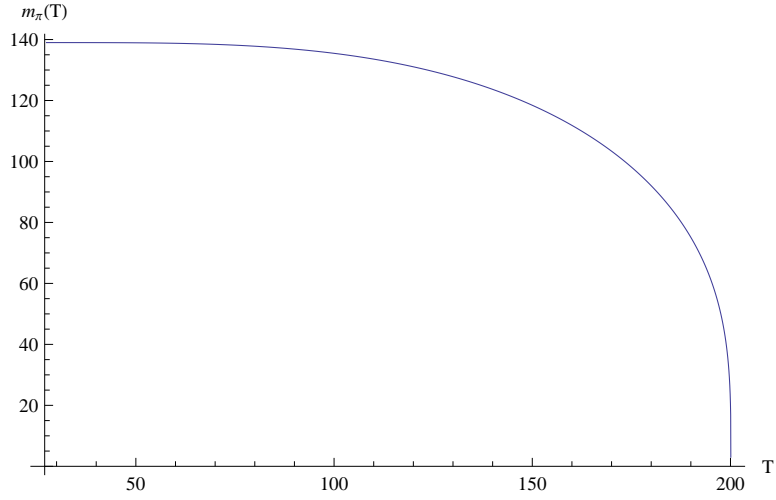}\\
  \caption{Temperature dependence of the pion mass}\label{mpi1}
\end{center}
\end{figure}

The leading order approximation of the pseudoscalar meson mass
$m_{\pi^{0,\pm}}$ with respect to current quark mass $m$ are
expressed in Eq. (\ref{pscalar}), which is shown in
Fig.(\ref{mpi1}).

Let us now turn to thermodynamic property of the pseudoscalar meson mass
Eq.(\ref{pscalar}):
\begin{eqnarray}
m_{\pi^{0,\pm}}^2(T) &\simeq&
\frac{4\mu_P^3(T)}{f_\pi^2(T)}m = \frac{4 v_o(T) \mu_f^2(T)
}{f_\pi^2(T)}m = \frac{\mu_f^2(T)}{\bar{\lambda}(T)v_o(T)}m
\end{eqnarray}
which explicitly shows that when keeping the mass scale $\mu_f^2$ to be a temperature-independent
constant, the thermodynamic mass of the pseudoscalar meson becomes divergent near the critical temperature $T_c$ as $v_o(T_c) =0$, which is obviously contrary to our intuition. This is a manifest reason why we should make an assumption for the temperature dependence of
$\mu_f^2(T)$ given in Eq.(\ref{mu_f}), which can lead to the expected thermodynamic
behavior for the pseudoscalar meson mass near the critical point
\begin{eqnarray}
m_{\pi^{0,\pm}}^2(T) = \frac{\gamma_o v_o(T)}{\bar{\lambda}(T)} m,
\end{eqnarray}
which is shown in Fig.(\ref{mpi1}).

According to the above derivation, we see that the phase transition
of chiral symmetry restoration is second order in our simple model.
Thus, it is natural to determine the critical behavior of all the
quantities discussed previously. Now we would like to find the the
scaling behavior of the vacuum expectation value $v_o(T)$ near the
critical point. By expanding our gap equation Eq.(\ref{gap eqn})
around the critical temperature $T_c$ with respect to the small
value of VEV $v_o(T)^2$ up to the order of $v_o(T)^2$, we can
obtain:
\begin{eqnarray}
\frac{1}{6}N_c(T_c^2-T^2)-C v_o(T)^2=0,
\end{eqnarray}
where $C =
\frac{N_c}{8\pi^2}\gamma(0,\frac{\mu_s^2}{M_c^2})+\frac{1}{4}-\gamma_o$
and $\gamma(s,x)\equiv \int^x_0t^{s-1}e^{-t}dt$ is the lower
incomplete gamma function. Given above equation, we can easily
obtain:
\begin{eqnarray}
v_o = \sqrt{\frac{N_c}{6C}}(T_c^2-T^2)^{\frac{1}{2}}\propto
(T_c-T)^{\frac{1}{2}}.
\end{eqnarray}
Thus, the critical dimension is $\beta=0.5$. Other quantities such
as $f_\pi(T)$, $m_{\pi^{0,\pm}}(T)$ and $(-\<\bar{q}q\>(T))^{1/3}$
all have the same scaling behavior. Such a critical behavior is not
accidental,  which can actually be understood from the fact that
near the critical temperature all these quantities are proportional
to the VEV $v_o(T)$ with $\bar{\lambda}(T)$ keeping fixed to
$\bar{\lambda}(T_c)\neq 0$.

\section{Conclusions and Remarks}

In this paper, we have extended the chiral dynamical model to the chiral thermodynamic model by adopting the CTPGF approach. The resulting effective chiral Lagrangian for the composite meson fields at finite temperature is similar to the one of the CDM, but all the couplings and mass scales become temperature dependent. We have discussed in detail the finite temperature behavior of CTDM. Much attention has been paid to the thermodynamic
chiral symmetry breaking and its restoration at finite temperature.
After fixing the free parameters in the effective chiral Lagrangian,
we have determined the critical temperature for the chiral symmetry
restoration, its value has been found to be around $T_c \simeq 200$ MeV which is consistent with other
predictions based on the NJL model\cite{Hatsuda1994,Klevansky1992,Alkofer1996,Buballa2005}.
We have also explicitly presented the thermodynamic behavior of several interesting quantities which include the vacuum expectation value VEV $v_o(T)$, the pion decay constant $f_\pi(T)$, the quark condensate
$\<\bar{q}q\>(T)$ and the pseudoscalar meson mass $m_{\pi^{0,\pm}}(T)$, they all display the property of the
chiral symmetry restoration at the critical temperature $T_c$. From the numerical calculations, we have shown that they all have the same
scaling behavior near the critical point. It is interesting to note that the mass scale $\mu_f$ for the four quark interaction in the NJL model should be temperature dependent at finite temperature as expected from the gluon thermodynamics, its thermodynamic behavior near the critical point is required to be same as the one of the chiral symmetry breaking. As a consequence, we are led to the assumption that $\mu_f^2(T)=\gamma_o v_o^2(T)$ in order to yield the expected thermodynamic behavior of the pion meson mass and to avoid the divergent behavior near the critical point of phase transition. Finally, we would like to remark that as limited from our main purpose in the present paper we have only
considered two flavor quarks and ignored the important instanton effects and U(1)$_A$ anomalous effect, which prevents us to discuss some other interesting properties, such as the large strange quark mass effects and the anomalous
$U(1)_A$ symmetry restoration at finite temperature, we shall investigate those interesting effects elsewhere.

\vspace{1 cm}

\centerline{{\bf Acknowledgement}}

\vspace{20 pt}

The authors would like to thank L.X. Cui  and Y.B. Yang for useful discussions. This work was supported in part by the
National Science Foundation of China (NSFC) under Grant \#No.
10821504, 10975170 and the Project of Knowledge Innovation Program
(PKIP) of the Chinese Academy of Science.

\newpage
\appendix

\section{Brief Outline on Closed-Time-Path Green Function (CTPGF) Formalism}

The formalism used in zero-temperature quantum field theory is
suitable to describe observables (e.g. cross-sections) measured in
empty space-time, as particle interactions in an accelerator.
However, at high temperature, the environment has a non-negligible
density of matter which makes the assumption of zero-temperature field
theories inapplicable. Namely, under those circumstances, the methods of zero-temperature field
theories are not sufficient any more and should be replaced by others, which is
closer to thermodynamics where the background state is a thermal bath. Therefore we shall develop
quantum field theory with finite temperature which is extremely useful to
study all phenomena due to the collective effects, such as:
phase transitions, early universes, etc. There are several approaches
for the finite temperature field theories, in this appendix we
shall focus on the closed-time-path Green function (CTPGF) formalism which is simply applicable in our case.
The CTPGF formalism, developed by Schwinger~\cite{Schwinger1961} and
Keldysh~\cite{Keldysh1965}, has been used to solve lots of
interesting problems in statistical physics and condensed matter
theory~\cite{Chou1985}. It is generally believed that this technique
is quite efficient in investigating the nonequilibrium and finite
temperature dynamical systems, this is because such a formalism naturally
incorporates both the statistical and dynamical information\cite{Chou1985,Zhou1980}.
Excellent review articles~\cite{Chou1985,Das:2000ft} and monographs
\cite{Calzetta2008,Rammer2007} have described different aspects
of these issues. In this appendix, we will briefly outline the main method with
the Schwinger-Keldysh propagators and the universal Feynman
rules for the general theory.

Let us begin by the general discussion of statistical physics. A
dynamical system can be characterized by its Hamiltonian $H$ and a
statistical ensemble of this system in equilibrium at a finite
temperature $T=\frac{1}{\beta}$ (in units of Boltzmann constant) is
described in terms of a partition function
\begin{equation}
Z(\beta)=\Tr\rho(\beta)=\Tr e^{-\beta\mathcal {H}}
\end{equation}
Here $\rho(\beta)$ is known as the density matrix operator and
$\mathcal{H}$ can be thought of as the generalized Hamiltonian of
the system. For example, for a canonical ensemble in which the
system can only exchange energy with the heat bath, $\mathcal{H}$ is
defined as:
\begin{equation}
\mathcal{H}=H
\end{equation}
while for a grand canonical ensemble in which the system can not
only exchange energy with the heat bath but also exchange particles
with the reservoir, ${\mathcal H}$ is taken as
\begin{equation}
\mathcal{H}=H-\mu N
\end{equation}
where $\mu$ is the chemical potential and $N$ represents the
particle number operator.

A observable in a statistical ensemble is the ensemble average for
any operator
\begin{equation}
\langle\mathcal{O}
\rangle_\beta=\frac{1}{Z(\beta)}\Tr\rho(\beta)\mathcal{O}
\end{equation}
Since the partition function and ensemble averages involve a
trace operation, this feature leads to the famous KMS
(Kubo-Martin-Schwinger) relation.
\begin{eqnarray}
\langle\mathcal{O}_1(t)\mathcal{O}_2(t^\prime)\rangle_\beta &=&
\frac{1}{Z(\beta)}\Tr
e^{-\beta\mathcal{H}}\mathcal{O}_1(t)\mathcal{O}_2(t^\prime)\nonumber\\
&=& \frac{1}{Z(\beta)}\Tr
e^{-\beta\mathcal{H}}\mathcal{O}_2(t^\prime) e^{-\beta\mathcal{H}}
\mathcal{O}_1(t) e^{\beta\mathcal{H}}\nonumber\\
&=& \frac{1}{Z(\beta)}\Tr
e^{-\beta\mathcal{H}}\mathcal{O}_2(t^\prime)
\mathcal{O}_1(t-i\beta)\nonumber\\
&=& \langle
\mathcal{O}_2(t^\prime)\mathcal{O}_1(t-i\beta)\rangle_\beta
\end{eqnarray}
Note that KMS relation only rely on the trace operation and does not
depend on any periodicity property of operators along the
temperature (imaginary time) interval.

Now it is easily seen that the operator $e^{-\beta\mathcal{H}}$ in
the definition of the partition function is very similar to the time
evolution operator in the imaginary time axis
$e^{-i(-i\beta)\mathcal{H}}$ \cite{Bloch1958}. We promote this
similarity and analytically extend the time variable to the complex
plane. So the operator $e^{-\beta\mathcal{H}}$ would live on the
line interval which is parallel to the negative imaginary time-axis,
with its length $\beta$. The analogy implies that we can define our
theory on some certain contour on the complex t-plane.

The contour should satisfy the following several requirement: (i) The two endpoints of the contour must be fixed in an interval in a
line parallel to the imaginary axis with its length $\beta$. The
stating point A (corresponding to time $t_i$) and the ending point
B(corresponding to $t_f=t_i-i\beta$) are identified, and one
requires that $\mathcal{O}|_B=\mathcal{O}|_A$ if $\mathcal{O}$ is
bosonic and and $\mathcal{O}|_B=-\mathcal{O}|_A$ if $O$ is
fermionic; (ii) For a system whose spectrum of the Hamiltonian is semi-positive
(at least bounded below due to the stability of the system), the
contour needs to have a monotonically decreasing or constant
imaginary part for the reason of  the convergence of the complete
partition function.The same result can also be obtained by analyzing
the convergence of the two-point Green
function.\cite{Quiros:1999jp}; (iii) The contour needs to pass the whole real axis of t-plane on which
the field operators $\mathcal{O}_i(t)$ are defined (Real-time
Formalism. Otherwise, like the imaginary time formalism, the field
operators need to be analytically extended to the imaginary axis
first).

\begin{figure}[ht]
\begin{center}
  \includegraphics[scale=0.6]{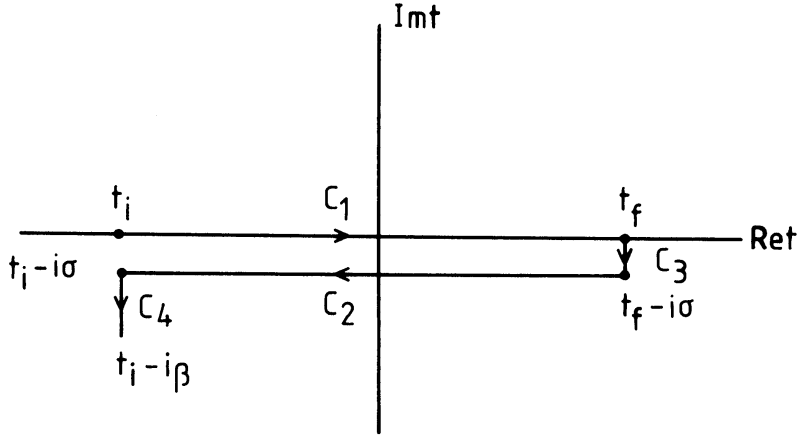}\\
  \caption{Contour used in the real time formalism}\label{realcontour}
\end{center}
\end{figure}

The particular family of such real time contours is depicted in
Fig.~\ref{realcontour}
where the contour ${\cal C}$ is ${\displaystyle {\cal C}=C_1\bigcup
C_2 \bigcup C_3 \bigcup C_4}$. The contour $C_1$ goes from the initial time
$t_i$ to the final time $t_f$, $C_3$ from $t_f$ to $t_f-i\sigma$,
with $0\le\sigma\le\beta$, $C_2$ from $t_f-i\sigma$ to
$t_i-i\sigma$, and $C_4$ from $t_i-i\sigma$ to $t_i-i\beta$.
Different choices of $\sigma$ lead to an equivalence class of
quantum field theories at finite temperature. Our preferred choice
is the Schwinger-Keldysh one with $\sigma=0$.

With this specific contour, the action of a field configuration is
the sum of contributions from the three parts,
\begin{equation}\label{action}
S = \int\limits_{\cal C}\!dt\, L(t) =
  \int\limits_\ti^\tf\!dt\, L(t)
  -\int\limits_\ti^\tf\!dt\, L(t)
  -i\!\int\limits_0^\beta\!d\tau\, L(\ti - i\tau)\,
\end{equation}
where \begin{equation}
  L(t) = \int\!d\vec{x}\, \mathcal{L}[\phi(t,\vec{x})]\,,
\end{equation}
and $\mathcal{L}$ is the Lagrangian density. In the following we
will take the theory of a scalar field $\phi(x)$ as an example.
However, in the limit $t_i\to -\infty$ and $t_f\to\infty$, it can be
shown that the third branch gets decoupled from the other two (the
factors in the propagators connecting such branches are
asymptotically damped). Consequently, in this limit, we are
effectively dealing with two branches leading to the name "closed
time path formalism". In this contour, then, the time integration
has to be thought of as
\begin{equation}\label{contourC}
\int_{\cal C} dt = \int^\infty_{-\infty} d t_+
-\int^{\infty}_{-\infty} dt_-
\end{equation}
where the relative negative sign arises because time is decreasing
in the second branch of the time contour.

The advantage of introducing the contour ${\cal C}$ is that one can
introduce the sources coupled to the field $\phi$ which is not
vanishing on the two Minkowski parts of the contour. This procedure
would give us the generating functional
\begin{equation}\label{gen-func}
  Z[J_1,J_2] = \int\!{\cal D}\phi\,\exp\left(iS
  +i\!\int\limits_{-\infty}^\infty\!dt_+\!\int\!d\vec{x}\,J_1(x)\phi_1(x)
  -i\!\int\limits_{-\infty}^\infty\!dt_-\!\int\!d\vec{x}\,J_2(x)\phi_2(x)
  \right) \ .
\end{equation}
Here $J_{1,2}$ and $\phi_{1,2}$ are the sources and fields on the
two Minkowski parts of the contour, i.e.,
\begin{subequations}
\begin{eqnarray}
  J_1(t,\vec{x}) = J(t_+,\vec{x})\,, & \qquad & \phi_1(t,\vec{x}) = \phi(t_+,\vec{x})\,,\\
  J_2(t,\vec{x}) = J(t_-,\vec{x})\,, & \qquad &
    \phi_2(t,\vec{x}) = \phi(t_-,\vec{x})\,.
\end{eqnarray}
\end{subequations}
By taking second variations of $Z$ with respect to the source $\phi$
one finds the Schwinger-Keldysh propagator
\begin{equation}
  iG_{ab}(x-y) =
  \frac 1{i^2}\,\frac{\delta^2\ln Z[J_1,J_2]}
  {\delta J_a(x)\,\delta J_b(y)}
   = i \left(\begin{array}{cc} G_{11} & -G_{12}\\-G_{21} & G_{22}
  \end{array}\right) \ .
\end{equation}
In the operator formalism, the Schwinger-Keldysh propagator
corresponds to the contour-ordered correlation function. In the single
time representation\cite{Chou1985}, this means:
\begin{equation}\label{Gab-oper}
\begin{split}
  iG_{11}(t,\x) = \< T \phi_1(t,\x)\phi_1(0)\>_{\beta}\,,\qquad &
  iG_{12}(t,\x) = \< \phi_2(0)\phi_1(t,\x)\>_{\beta}\,,\\
  iG_{21}(t,\x) = \< \phi_2(t,\x)\phi_1(0)\>_{\beta}\,,\qquad &
  iG_{22}(t,\x) = \< \bar T \phi_2(t,\x)\phi_2(0)\>_{\beta}\,.
\end{split}
\end{equation}
where $T$ ($\bar{T}$) denotes normal (reversed) time ordering, and
\begin{subequations}
\begin{eqnarray}
  \phi_1(t,\x) &=& e^{iHt-i\P\cdot\x} \phi(0) e^{-iHt+i\P\cdot\x}\,,\\
  \phi_2(t,\x) &=& e^{iH(t-i\sigma)-i\P\cdot\x} \phi(0)
     e^{-iH(t-i\sigma)+i\P\cdot\x}\,.
\end{eqnarray}
\end{subequations}

Let us now consider the free real scalar theory. If one goes to the
momentum space and by inserting the complete set of states into the
definitions (\ref{Gab-oper}), one finds the explicit form of the
previously defined Schwinger-Keldysh propagator.
\begin{eqnarray}\label{SKprop}
  iG_{11}(k) &=& \frac{i}{k^2-m^2+i\epsilon}+2\pi
  n_B(\omega)\delta(k^2-m^2)\,,\qquad \omega\equiv|k_0|\\
  iG_{12}(k) &=& 2\pi
  [n_B(\omega)+\theta(-k_0)]\delta(k^2-m^2)\,,\\
 iG_{21}(k) &=& 2\pi
  [n_B(\omega)+\theta(k_0)]\delta(k^2-m^2)\,,\\
 iG_{22}(k) &=& -\frac{i}{k^2-m^2+i\epsilon}+2\pi
  n_B(\omega)\delta(k^2-m^2)\,.
\end{eqnarray}
Or in the matrix form:
\begin{eqnarray}
iG_\beta(k) &=& \left(\begin{array}{cc}\frac{i}{k^2-m^2+i\epsilon} &
{2\pi\theta(-k_0)\delta(k^2-m^2)}
\\ {2\pi\theta(k_0)\delta(k^2-m^2)} & -\frac{i}{k^2-m^2-i\epsilon}\end{array}\right)\nonumber\\
&+& 2\pi
  n_B(\omega)\delta(k^2-m^2)\left(\begin{array}{cc} 1 & 1 \\ 1 & 1
\end{array}\right)
\end{eqnarray}
where $n_B(\omega)\equiv\frac{1}{e^{\beta\omega}-1}$ stands for the
Bose-Einstein distribution function. Note that the propagator is a
$2\times2$ matrix, a consequence of the doubling of the degrees of
freedom. However, the propagators (12), (21) and (22) are unphysical
since at least one of their time arguments is on the negative
branch. They are required for the consistency of the theory. The
only physical propagator is the (11) component shown in
Eq.(\ref{SKprop}).

For perturbative calculations we need to know the complete Feynman
rules besides of the propagators. From the generating functional
Eq.(\ref{gen-func}) and the action Eq.(\ref{action}) defined on
contour ${\cal C}$ Eq.(\ref{contourC}), we see that the complete
theory contains two types of vertices- type-1 for the original
fields $\phi_1(x)$ while type-2 for the doubled fields $\phi_2(x)$.
The vertices for the partner fields will have a relative negative
sign corresponding to the original vertices, because time is
decreasing in the negative branch. The four possible propagators,
(11), (12), (21) and (22) defined above connect them. All of them
have to be considered for the consistency of the theory. The golden
rule is that: physical legs must always be attached to type 1
vertices\cite{Quiros:1999jp}. For other Feynman rules, including the
integration measure, the symmetry factors involved in Feynman
diagrams, the topology of the Feynman diagrams, etc. are all the
same as the zero-temperature field theory.

For the application to the present paper, we also need to know the
Schwinger-Keldysh propagator for fermions as the quark fields here
are represented as the chiral fermion fields
\begin{eqnarray}
iS_\beta(k) &=&(k\sla+m)[
\left(\begin{array}{cc}\frac{i}{k^2-m^2+i\epsilon} &
{2\pi\theta(-k_0)\delta(k^2-m^2)}
\\ {2\pi\theta(k_0)\delta(k^2-m^2)} & -\frac{i}{k^2-m^2-i\epsilon}\end{array}\right)\nonumber\\
&& - 2\pi
  n_F(\omega)\delta(k^2-m^2)\left(\begin{array}{cc} 1 & 1 \\ 1 & 1
\end{array}\right)]
\end{eqnarray}
where $n_F(\omega)\equiv\frac{1}{e^{\beta\omega}+1}$ stands for the
Fermi-Dirac distribution function.

When transforming into the Euclidean spacetime the Schwinger-Keldysh
propagator defined above becomes
\begin{eqnarray}
iS_{E\beta}(k) &=&(-i)(k\sla_E+m)[
\left(\begin{array}{cc}\frac{1}{k_E^2+m^2} & {2\pi
i\theta(-k_{E4})\delta(k_E^2+m^2)}
\\ {2\pi i\theta(k_{E4})\delta(k_E^2+m^2)} & -\frac{1}{k_E^2+m^2}\end{array}\right)\nonumber\\
&& - 2\pi i
  n_F(\omega)\delta(k_E^2+m^2)\left(\begin{array}{cc} 1 & 1 \\ 1 & 1
\end{array}\right)]
\end{eqnarray}
In Sec. 3 the factor $-i$ is canceled by the factor $i$ in the
integration measure transformation $d^4k\to i d^4 k_E$.
\section{Derivation of Chiral Dynamical Model in the Chiral Rotated
Basis}\label{appendix}
In this appendix, we will derive the
effective chiral Lagrangian for mesons in the so-called chiral
\textquotedblleft Rotated Basis"\cite{Espriu:1989ff}. Although the
obtained effective chiral Lagrangian will not change, it is more
transparent to see the chiral symmetries and their spontaneous
breaking in this derivation. Let us begin with the effective
Lagrangian
\begin{equation}\label{Lag3}
{\cal L}^q_{eff}(q,{\bar q}) = \bar{q}\gamma^\mu i\partial_\mu
q+\bar{q}_L\gamma_\mu {\cal A}^\mu_L q_L+\bar{q}_R\gamma_\mu {\cal
A}_R^\mu q_R -[\bar{q}_L(\Phi-M)q_R+h.c.]
\end{equation}
where the auxiliary meson fields $\Phi(x)$ is defined as in
Eq.(\ref{meson_def})
\begin{eqnarray}
&& \Phi(x)\equiv \xi_L(x)\phi(x)\xi_R^\dagger(x),\qquad U(x)\equiv
\xi_L(x)\xi_R^\dagger(x)=\xi_L^2(x)=e^{i\frac{2\Pi(x)}{f}}\nonumber\\
&& \phi^\dagger(x)=\phi(x)=\sum^{3}_{a=0}\phi^a(x)T^{a},\qquad
\Pi^\dagger(x)=\Pi(x)=\sum^{3}_{a=0}\Pi^a(x)T^a,
\end{eqnarray}
where $\Pi(x)$ and $\phi(x)$ represent the pseudoscalar and scalar
mesons respectively. Note that except for the mass term or the
source term, the Lagrangian is invariant under the transformation of
the local $U(2)_L\times U(2)_R$ chiral symmetry:
\begin{eqnarray}
&&q_L(x)\to g_L(x)q_L(x),\quad q_R(x)\to g_R(x)q_R(x);\quad
\Phi(x)\to
g_L(x)\Phi(x)g_R^\dagger(x),\nonumber\\
&& {\cal A}_{L\mu}\to g^\dagger_L {\cal A}_{L\mu}g_L(x) -i g^\dagger_L\partial_\mu
g_L(x), \quad {\cal A}_{R\mu} \to g^\dagger_R {\cal A}_{R\mu}g_R(x) -i
g^\dagger_R\partial_\mu g_R(x),
\end{eqnarray}
The transformation for $\Phi(x)$ can also be written in terms of the fields
$\phi(x)$ and $\xi_L(x)$ as:
\begin{eqnarray}
\phi(x) \to h(x)\phi(x)h^\dagger(x),\quad \xi_L(x)\to
g_L(x)\xi_L(x)h^\dagger(x)=h(x)\xi_L(x)g^\dagger_R(x).
\end{eqnarray}
Let us now introduce new quark fields, which is referred to the chiral
\textquotedblleft Rotated Basis" in\cite{Espriu:1989ff}.
\begin{eqnarray}
q_L = \xi_L Q_L,&\quad& \bar{q}_L=\bar{Q}_L
\xi_L^\dagger,\nonumber\\
q_R = \xi_L^\dagger Q_R,&\quad& \bar{q}_R=\bar{Q}_R\xi_L.
\end{eqnarray}
With this new quark basis, we can rewrite the Lagrangian
Eq.(\ref{Lag3}) in the following form
\begin{eqnarray}\label{Lag4}
{\cal L}^Q_{eff}(Q,{\bar Q}) = \bar{Q}\gamma^\mu i\partial_\mu
Q+\bar{Q}_L\gamma^\mu L_\mu Q_L+\bar{Q}_R\gamma^\mu R_\mu Q_R
-[\bar{Q}_L(\phi-{\cal M})Q_R+h.c.],
\end{eqnarray}
where the fields $L_\mu, ~R_\mu$ and ${\cal M}$ are defined as
\begin{eqnarray}\label{transform}
L_\mu &\equiv& \xi_L^\dagger{\cal
A}_{L\mu}\xi_L+i\xi_L^\dagger\partial_\mu\xi_L,\quad R_\mu \equiv
\xi_L{\cal
A}_{R\mu}\xi^\dagger_L+i\xi_L\partial_\mu\xi^\dagger_L, \nonumber\\
{\cal M} &\equiv& \xi_L^\dagger M \xi_L^\dagger, \quad {\cal
M}^\dagger \equiv \xi_L M^\dagger \xi_L.
\end{eqnarray}
In the above \textquotedblleft rotated basis", the quark fields
$Q_{L(R)}(x)$ transform only under the diagonal $U_V(2)$ symmetry:
\begin{eqnarray}
Q_L(x) \to h(x) Q_L(x), \quad Q_R(x) \to h(x) Q_R(x).
\end{eqnarray}
Thus, the quark fields $Q_{L(R)}(x)$ are much like the
\textquotedblleft constituent quark" defined in the nonrelativistic
quark model \cite{Manohar:1983md}. When the chiral symmetry is
spontaneous breaking and the meson field $\phi(x)$ acquires the
vacuum expectation value (VEV) $<\phi(x)>=V$, $Q_{L(R)}$ will obtain
a mass term $\bar{Q}_L (V-{\cal M}) Q_R + h.c.$. In the case where
each quark flavor possesses the universal current mass $m$, the VEV
matrix is diagonal $V=v\cdot I$, here $I$ is the identity matrix in
the flavor space. The mass term is $(v-m)\bar{Q}_L Q_R+h.c.$, namely the mass of quarks $Q(x)_{L(R)}$ is the dynamical quark mass
$\bar{m}= v-m$ defined in Sec.\ref{sec2}.

Note that the Lagrangian Eq.(\ref{Lag4}) has the same structure as
the original one Eq.(\ref{Lag3}) except for the definition of the
mass and gauge fields. Thus, we may expect that the effective chiral
Lagrangian for the meson fields has the same structure as
Eq.(\ref{Lag_meson}). Indeed, by integrating over the quark fields
following the procedure from Eq.(\ref{Lag2}) to Eq.(\ref{ChEF3}), we
obtain
\begin{eqnarray}\label{ChEF4}
S^M_{E Re} = \frac{N_c}{2}\int d^4 x_E \int \frac{d^4k}{(2\pi)^4}
tr_{SF} \ln(k^2+ \Delta^\prime_E)-\ln Z_0,
\end{eqnarray}
where ${\Delta}^\prime_E$ above is defined as
\begin{eqnarray}
\Delta^\prime_E &=& \hat{\phi}\hat{\phi}^\dagger P_R +
\hat{\phi}^\dagger \hat{\phi} P_L - i\gamma\cdot D^\prime_E \phi P_L
-
i\gamma\cdot D^\prime_E\phi^\dagger P_R\nonumber\\
&& -\sigma_{\mu\nu}{\cal F}^\prime_{R\mu\nu} P_L - \sigma_{\mu\nu}
{\cal F}^\prime_{L\mu\nu} P_R +(i
D^\prime_{E\mu})(iD^\prime_{E\mu})+2 k\cdot(i D^\prime_E),
\end{eqnarray}
and
\begin{eqnarray}
i D^\prime_{E\mu} \phi &=& i\partial_\mu \phi + L_\mu \phi - \phi
R_\mu,\\
i D^\prime_{E\mu} &=& i\partial_\mu + R_\mu P_L + L_\mu P_R,\\
\hat{\phi} &\equiv& \phi-{\cal M}.
\end{eqnarray}
In order to derive the effective action for meson field, we redefine
$\Delta^{k\prime}_E\equiv k^2+ \Delta^\prime_E$ to the following two
terms:
\begin{eqnarray}
\Delta^{k\prime}_E \equiv k^2+ \Delta^\prime_E = \Delta_0
+\tilde{\Delta}^\prime_E,
\end{eqnarray}
with
\begin{eqnarray}
\Delta_0 &=& k^2+\bar{M}^2,\nonumber\\
\tilde{\Delta}^\prime_E &=& [(\hat{\phi}\hat{\phi}^\dagger-\bar{\cal
M}\bar{\cal M}^\dagger) P_R +(\hat{\phi}^\dagger\hat{\phi}-\bar{\cal
M}^\dagger\bar{\cal M})P_L-i\gamma\cdot D_E^\prime \phi P_L
-i\gamma\cdot D^\prime_E\phi^\dagger P_R
\nonumber\\
&& -\sigma_{\mu\nu} {\cal
F}^\prime_{R\mu\nu}P_L-\sigma_{\mu\nu}{\cal F}^\prime_{L\mu\nu} P_R
+ (i D^\prime_{E\mu})(i D^\prime_{E\mu}) +
2k\cdot(iD_E^\prime)] \nonumber\\
&& +(\bar{\cal M}\bar{\cal M}^\dagger - \bar{M}^2)P_R + (\bar{\cal
M}^\dagger\bar{\cal M}-\bar{M}^2)P_L,
\end{eqnarray}
where $\bar{M}$ is supposed vacuum expectation values (VEVs) of
$\hat{\Phi}$, i.e., $<\hat{\Phi}> = \bar{M}$ which is real and
$\bar{\cal M} \equiv \xi^\dagger_L\bar{M}\xi^\dagger_L$. As we will
see that the terms in the third line of the definition of
$\tilde{\Delta}^\prime_E$ is crucial to prove the equivalence of the
obtained effective action for mesons in this rotated basis to the
one given in Eq.(\ref{Lag_meson}).

Now if we regard $\tilde{\Delta}_E^\prime$ as the perturbation and
take $Z_0=(\det \Delta_0)^{\frac{1}{2}}$ as before, we can expand
the effective action Eq.(\ref{ChEF4}) according to
$\tilde{\Delta}_E^\prime$:
\begin{eqnarray}\label{Lag5}
S^M_{E Re} &=& \frac{N_c}{2} \int d^4 x_E \int\frac{d^4k}{(2\pi)^4}
tr_{SF}[\ln(\Delta_0+\tilde{\Delta}^\prime_E)-\ln\Delta_0]\nonumber\\
&=& \frac{N_c}{2} \int d^4 x_E \int\frac{d^4k}{(2\pi)^4}
tr_{SF}\ln(1+\frac{1}{\Delta_0}\tilde{\Delta}^\prime_E)\nonumber\\
&=& \frac{N_c}{2} \int d^4 x_E \int\frac{d^4k}{(2\pi)^4} tr_{SF}
\sum^\infty_{n=1}\frac{(-1)^{n+1}}{n}(\frac{1}{\Delta_0}\tilde{\Delta}^\prime_E)^n.
\end{eqnarray}
If we only keep the leading two terms in the expansion as in the
previous sections, we can obtain:
\begin{eqnarray}\label{ChEF5}
S^M_{E Re} &\approx& \frac{N_c}{2}\int d^4 x_E \int \frac{d^4
k}{(2\pi)^4}
tr_{SF}[\frac{1}{\Delta_0}\tilde{\Delta}^\prime_E-\frac{1}{2}\frac{1}{\Delta_0^2}(\tilde{\Delta}^\prime_E)^2]\nonumber\\
&=& \frac{N_c}{16\pi^2}\int d^4 x_E tr_F
\{M_c^2L_2[(\hat{\phi}\hat{\phi}^\dagger-\bar{\cal M}\bar{\cal
M}^\dagger)+ (\hat{\phi}^\dagger\hat{\phi}-\bar{\cal
M}^\dagger\bar{\cal M})]\nonumber\\
&&-\frac{1}{2} L_0 [D^\prime_E\hat{\phi}\cdot
D^\prime_E\hat{\phi}^\dagger+ D^\prime_E\hat{\phi}^\dagger\cdot
D^\prime_E\hat{\phi}+ (\hat{\phi}\hat{\phi}^\dagger-\bar{\cal
M}\bar{\cal M}^\dagger)^2+(\hat{\phi}^\dagger\hat{\phi}-\bar{\cal
M}^\dagger\bar{\cal M})^2]\nonumber\\
&& +M_c^2L_2[(\bar{\cal M}\bar{\cal M}^\dagger-\bar{M}^2)+(\bar{\cal
M}^\dagger\bar{\cal
M}-\bar{M}^2)]\nonumber\\
&& -\frac{1}{2}L_0[(\hat{\phi}\hat{\phi}^\dagger-\bar{\cal
M}\bar{\cal M}^\dagger)(\bar{\cal M}\bar{\cal
M}^\dagger-\bar{M}^2)+(\hat{\phi}^\dagger\hat{\phi}-\bar{\cal
M}^\dagger\bar{\cal M})(\bar{\cal M}^\dagger\bar{\cal
M}-\bar{M}^2)\nonumber\\
&&+(\bar{\cal M}\bar{\cal
M}^\dagger-\bar{M}^2)(\hat{\phi}\hat{\phi}^\dagger-\bar{\cal
M}\bar{\cal M}^\dagger)+(\bar{\cal M}^\dagger\bar{\cal
M}-\bar{M}^2)(\hat{\phi}^\dagger\hat{\phi}-\bar{\cal
M}^\dagger\bar{\cal M})\nonumber\\
&& (\bar{\cal M}\bar{\cal M}^\dagger-\bar{M}^2)^2+(\bar{\cal
M}^\dagger\bar{\cal M}-\bar{M}^2)^2]\},
\end{eqnarray}
where the matrix $L_0$ and $L_2$ are defined in Eq.(\ref{intg1}). It
is easy to see that the last four lines of terms vanish if the
different flavors of quarks have the same current mass which leads
to the same vacuum expectation value for each flavor:
\begin{eqnarray}
\bar{\cal M}\bar{\cal M}^\dagger-\bar{M}^2 &=& (\xi_L^\dagger
\bar{M}
\xi_L^\dagger)(\xi_L\bar{M}\xi_L)-\bar{M}^2 \nonumber\\
&=& \xi_L^\dagger \bar{M}^2\xi_L-\bar{M}^2=0.
\end{eqnarray}
The last line is valid since the mass matrix $\bar{M}$ is diagonal
with the same eigenvalues and $\bar{M}$ can commute with the SU(2)
matrix $\xi_L$.

Note also that in the case of 2 flavors with the universal current
quark mass, after the chiral symmetry $SU_L(2)\times SU_R(2)$ is
broken to $SU_V(2)$ and the field $\phi(x)$ acquires vacuum
expectation value (VEV) $V=v\cdot I$, we have:
\begin{eqnarray}
iD_{E\mu}^\prime \hat{\phi} &\approx& (v-m)[L_\mu-R_\mu]\nonumber\\
&=& (v-m)i[\xi_L^\dagger(\partial_\mu-i{\cal
A}_{L\mu})\xi_L-\xi_L(\partial_\mu-i{\cal
A}_{R\mu})\xi^\dagger_L]\nonumber\\
&=& (v-m)i\xi_L^\dagger(D_{E\mu} U)\xi_L^\dagger =
-(v-m)i\xi_L(D_{E\mu} U^\dagger)\xi_L.
\end{eqnarray}
In order to prove the last two equalities, we have to use the
definition of $U\equiv \xi_L^2$ and the identity $\partial_\mu \xi_L
=-\xi_L(\partial_\mu\xi_L^\dagger)\xi_L$. Therefore, from the
kinetic terms for $\phi(x)$ in Eq.(\ref{ChEF5}), we can obtain the
kinetic term for the pseudoscalar meson field $U\equiv
e^{i2\Pi(x)/f_\pi}$:
\begin{eqnarray}
-\frac{N_c(v-m)}{16\pi^2}\int d^4 x_E tr_E [L_0(D_{E\mu}U)(D_{E\mu
}U^\dagger)].
\end{eqnarray}
However, when the chiral symmetry is restored, as discussed in the
context of CDTM, this term will disappear as the VEV of $\phi(x)$
vanishes. Thus, the discussion in this \textquotedblleft rotated
basis" gives a more transparent picture of Goldstone boson character
of the pseudoscalar mesons.

Next we would like to prove the equivalence between the effective
chiral Lagrangians Eq.(\ref{Lag_meson}) and Eq.(\ref{ChEF5}).
According to the definition of the transformations
Eq.(\ref{transform}), we can easily obtain the following relations:
\begin{eqnarray}
D^\prime_{E\mu} \hat{\phi} = \xi^\dagger_L(D_{E\mu}
\hat{\Phi})\xi^\dagger_L \quad D^\prime_{E\mu} \hat{\phi}^\dagger =
\xi_L(D_{E\mu} \hat{\Phi}^\dagger)\xi_L,
\end{eqnarray}
where $D_{E\mu}\hat{\Phi}$ and $D_{E\mu}\hat{\Phi}^\dagger$ are
defined as in Eq.(\ref{partial}). Thus, the first two lines of terms
in Eq.(\ref{ChEF5}) can be written in a form with respect to $\Phi$:
\begin{eqnarray}\label{ChEF6}
S^M_{E Re} &=& \frac{N_c}{2}\int d^4 x_E \text{tr}_F \{M_c^2
L_2[\xi^\dagger_L(\hat{\Phi}\hat{\Phi}^\dagger-\bar{M}^2)\xi_L
+\xi_L(\hat{\Phi}^\dagger\hat{\Phi}-\bar{M}^2)\xi_L^\dagger]\nonumber\\
&& -\frac{1}{2}L_0[\xi_L^\dagger(D_E\hat{\Phi}\cdot
D_E\hat{\Phi}^\dagger)\xi_L + \xi_L(D_E\hat{\Phi}^\dagger\cdot
D_E\hat{\Phi})\xi_L^\dagger \nonumber\\
&& + \xi_L^\dagger(\hat{\Phi} \hat{\Phi}^\dagger-\bar{M}^2)^2\xi_L +
\xi_L(\hat{\Phi}^\dagger
\hat{\Phi}-\bar{M}^2)^2\xi_L^\dagger]\}\nonumber\\
&=& \frac{N_c}{2}\int d^4 x_E \text{tr}_F \{M_c^2 [(\xi_L
L_2\xi^\dagger_L)(\hat{\Phi}\hat{\Phi}^\dagger-\bar{M}^2)
+(\xi_L^\dagger L_2\xi_L)(\hat{\Phi}^\dagger\hat{\Phi}-\bar{M}^2)]\nonumber\\
&& -\frac{1}{2}\{(\xi_LL_0\xi_L^\dagger)[D_E\hat{\Phi}\cdot
D_E\hat{\Phi}^\dagger+(\hat{\Phi} \hat{\Phi}^\dagger-\bar{M}^2)^2]
\nonumber\\ &&+ (\xi_L^\dagger L_0\xi_L)[D_E\hat{\Phi}^\dagger\cdot
D_E\hat{\Phi}+(\hat{\Phi}^\dagger \hat{\Phi}-\bar{M}^2)^2]\}\}.
\end{eqnarray}
In general the matrices $\xi_L$ and $L_0$ ($L_2$) do not commute
with each other when different flavors do not have the same current
masses. So the part of effective chiral Lagrangian shown in
Eq.(\ref{ChEF6}) is in general not equivalent to Eq.(\ref{Lag_meson}) by just comparing to the same truncated terms. In fact, 
by taking into account the higher-order terms in Eq.(\ref{Lag5}) and
the last four lines of terms in Eq.(\ref{ChEF5}), it is expected that
the extra terms would cancel the unwanted terms in
Eq.(\ref{ChEF6}) due to non-commutativity of $\xi_L$ and $L_0$
($L_2$). In our present case, as we only consider two
flavors with the same current quark mass, $\xi_L$ can commute with $L_0$ and $L_2$, there is no such extra terms.
Furthermore, as mentioned before, the last four lines of terms in
Eq.(\ref{ChEF5}) will vanish due to commutativity of $\xi_L$ and
$\bar{M}$. Thus, we arrive at the following effective chiral
Lagrangian:
\begin{eqnarray}
S^M_{E Re} &=& \frac{N_c}{16\pi^2}\int d^4 x_E tr_F
\{M_c^2L_2[(\hat{\Phi}\hat{\Phi}^\dagger-\bar{M}^2)+(\hat{\Phi}^\dagger\hat{\Phi}-\bar{M}^2)]\nonumber\\
&&-\frac{1}{2} L_0 [D_E\hat{\Phi}\cdot D_E\hat{\Phi}^\dagger+
D_E\hat{\Phi}^\dagger\cdot D_E\hat{\Phi}+
(\hat{\Phi}\hat{\Phi}^\dagger-\bar{M}^2)^2+(\hat{\Phi}^\dagger\hat{\Phi}-\bar{M}^2)^2]\},
\end{eqnarray}
which is exactly agree with Eq.(\ref{Lag_meson}) obtained in the
original unrotated basis.

Following the procedure in Sec.\ref{title}, we can derive the
effective chiral Lagrangian at finite temperature in the chiral
thermodynamic model (CTDM), which is:
\begin{eqnarray}\label{ChEF7}
S^M_{E Re} &\approx& \frac{N_c}{16\pi^2}\int d^4 x_E tr_F
\{M_c^2L_2(T)[(\hat{\phi}\hat{\phi}^\dagger-\bar{\cal M}\bar{\cal
M}^\dagger)+ (\hat{\phi}^\dagger\hat{\phi}-\bar{\cal
M}^\dagger\bar{\cal M})]\nonumber\\
&&-\frac{1}{2} L_0(T) [D^\prime_E\hat{\phi}\cdot
D^\prime_E\hat{\phi}^\dagger+ D^\prime_E\hat{\phi}^\dagger\cdot
D^\prime_E\hat{\phi}+ (\hat{\phi}\hat{\phi}^\dagger-\bar{\cal
M}\bar{\cal M}^\dagger)^2+(\hat{\phi}^\dagger\hat{\phi}-\bar{\cal
M}^\dagger\bar{\cal M})^2]\nonumber\\
&& +M_c^2L_2(T)[(\bar{\cal M}\bar{\cal
M}^\dagger-\bar{M}^2)+(\bar{\cal M}^\dagger\bar{\cal
M}-\bar{M}^2)]\nonumber\\
&& -\frac{1}{2}L_0(T)[(\hat{\phi}\hat{\phi}^\dagger-\bar{\cal
M}\bar{\cal M}^\dagger)(\bar{\cal M}\bar{\cal
M}^\dagger-\bar{M}^2)+(\hat{\phi}^\dagger\hat{\phi}-\bar{\cal
M}^\dagger\bar{\cal M})(\bar{\cal M}^\dagger\bar{\cal
M}-\bar{M}^2)\nonumber\\
&&+(\bar{\cal M}\bar{\cal
M}^\dagger-\bar{M}^2)(\hat{\phi}\hat{\phi}^\dagger-\bar{\cal
M}\bar{\cal M}^\dagger)+(\bar{\cal M}^\dagger\bar{\cal
M}-\bar{M}^2)(\hat{\phi}^\dagger\hat{\phi}-\bar{\cal
M}^\dagger\bar{\cal M})\nonumber\\
&& (\bar{\cal M}\bar{\cal M}^\dagger-\bar{M}^2)^2+(\bar{\cal
M}^\dagger\bar{\cal M}-\bar{M}^2)^2]\}.
\end{eqnarray}

By using the same argument as the one for the zero-temperature chiral
dynamical model (CDM) analyzed in the text, we can prove the equivalence between
Eq.(\ref{ChEF7}) and Eq.(\ref{ChEF8}). Thus, based on the equivalent
Lagrangians, the resulting physics, especially the spectrum of mesons, will
not be changed.


\end{document}